\documentclass{jfm} 
\usepackage{graphicx} 
\usepackage{natbib}
\usepackage{upmath}

\title[Surface granular flows]{Rheology of surface granular flows}

\author[Ashish V. Orpe and D. V. Khakhar]{A\ls S\ls H\ls I\ls S\ls
  H\ns V.\ns O\ls R\ls P\ls E\ns \and D.\ns V.\ns K\ls H\ls A\ls K\ls
  H\ls A\ls R\ls}

\affiliation{Department of Chemical Engineering, Indian Institute of
  Technology - Bombay, Powai, Mumbai 400076, India}

\begin{document}
\bibliographystyle{jfm}

\maketitle

\begin{abstract}
  Surface granular flow, comprising granular material flowing on the
  surface of a heap of the same material, occurs in several industrial
  and natural systems. The rheology of such flow is investigated by
  means of measurements of velocity and number density profiles in a
  quasi-two-dimensional rotating cylinder, half-filled with a model
  granular material - mono-size spherical stainless steel particles.
  The measurements are made at the center of the cylinder, where the
  flow is fully-developed, using streakline photography and image
  analysis . The stress profile is computed from the number density
  profile using a force balance which takes into account wall
  friction.  Mean velocity and root mean square (r.m.s.) velocity
  profiles are reported for different particle sizes and cylinder
  rotation speeds. The profiles for the mean velocity superimpose when
  distance is scaled by the particle diameter ($d$) and the velocity
  by a characteristic shear rate ($\dot{\gamma}_C =
  [g\sin(\beta_m-\beta_s)/d\cos\beta_s]^{1/2}$) and the particle diameter where
  $\beta_m$ is the static angle of friction and $\beta_s$ is the sliding angle
  of friction. The scaling is also found to work for the r.m.s.\ 
  velocity profiles. The mean velocity is found to decay exponentially
  with depth in the bed with a decay length of $\lambda=1.1d$. The r.m.s.\ 
  velocity shows similar behaviour but with $\lambda=1.7d$. The r.m.s.
  velocity profile shows two regimes: near the free surface the r.m.s.
  velocity is nearly constant and below a transition point it decays
  linearly with depth. The shear rate, obtained by numerical
  differentiation of the velocity profile, is not constant anywhere in
  the layer and shows a maximum which occurs at the same depth as the
  transition in the r.m.s. velocity profile. Above the transition
  point the velocity distributions are Gaussian and below the
  transition point the velocity distributions gradually approach a
  Poisson distribution. The shear stress increases roughly linearly
  with depth. However, the variation of the apparent viscosity ($\eta$)
  with r.m.s.\ velocity ($u$) shows a relatively sharp transition at
  the shear rate maximum, and in the region below this point the
  apparent viscosity varies as $\eta\sim u^{-1.5}$. The measurements
  indicate that the flow comprises two layers: an upper low viscosity
  layer with a nearly constant r.m.s.\ velocity and a lower layer of
  increasing viscosity with a decreasing r.m.s. velocity. The
  thickness of the upper layer depends on the local flow rate and is
  independent of particle diameter while the reverse if found to hold
  for lower layer thickness. The experimental data is compared to
  predictions of three models for granular flow.
 
\end{abstract}

\section{Introduction}
\label{sec:introduction}

Surface flows of granular materials comprise a shallow layer of particles flowing on a fixed bed of the same particles. Such flows are encountered in several industrial operations as well as in nature. Industrial examples appear in the processing of material in systems such as rotary kilns \citep{peray86} and tumbling mixers \citep{perry97}, in the formation of heaps for storage and in the feeding and discharge of hoppers and silos \citep{ned92}.  Natural examples are the formation of sand dunes \citep*{and02a,and02b,wang2004}, avalanches \citep*{hut95,hutt97} and transport of sediment in rivers \citep{boun02}.

Surface granular flows have one unique feature that differentiates them from other granular shear flows. The `boundary' between the flowing particles and the fixed bed is determined by the flow itself. A consequence is that interchange of particles between the flowing layer and the bed is possible, and the local surface angle and the local layer thickness vary with the local flow. Although it is reasonable to picture the flow in terms of two distinct regions (flowing layer and fixed bed) separated by a sharp interface, in fact the boundary is diffuse and the flow decays to zero exponentially with depth. Thus a homogeneous and responsive boundary exists at the base of the flow which is quite different in qualitative terms from a flow on an inclined rough surface.  Several studies have been carried out to understand the physics of surface granular flows and a review of previous work is given below.

Rotating cylinders and heaps have been the primary systems used for the experimental study of surface granular flows. In both systems, the major component of the velocity is parallel to the surface and the velocity profile is found to be broadly linear across the layer \citep{nak93,raj95,bon02a,jain02,longo02,tab03,orp04}. \citet{kom01} showed the decay of the velocity to be exponential with depth and this has been confirmed in several subsequent studies \citep{jain02,bon02a,longo02,tab03,orp04}.  The region of exponential decay is relatively small and there has been some focus on the velocity gradient corresponding to the linear part. Some studies show that the shear rate is nearly constant and independent of the local mass flow rate, particularly for low mass flow rates and near 2d systems \citep{raj95,kha01b,bon02a}. In other cases the shear rate is found to increase with flow rate \citep{orp01,jain02}. The bulk density is nearly constant across the layer and approaches close packing \citep{raj95,bon02a,jain02,orp04}, indicating the highly dissipative nature of the flow. 

The surface angle ($\beta$) is found to increase with mass flow rate in the layer \citep{kha01b,orp01}, however, in these and similar studies the measurements are made near a wall. Recently \citet{tab03} have shown a systematic effect of the wall on the surface angle. They found that the measured increase in angle with flow rate in heap flow experiments could be accounted for by considering the effect of wall friction in a force balance. The analysis yields   $\tan\beta=\mu_i+\mu_w\delta/b$, where $\delta$ is the layer thickness, $b$ is the gap width, $\mu_i$ is the effective coefficient of friction and $\mu_w$ the coefficient of friction beteween the particles and wall. Thus for sufficiently wide gaps ($\delta/b>>1$) a constant surface angle should be obtained. This was indeed observed by \citet{jop05} in a systematic experimental study of the effect of gap width on heap flow. They found that the gap width between side walls also affects the mean velocity and the depth of the flowing layer. As the gap width is increased the surface angle decreases, the mean velocity decreases and the depth of the flowing layer increases. The rate of increase of surface angle with flow rate reduces with gap width, and at the highest gap width considered (570 particle diameters) the surface angle is nearly independent of the flow rate. Down-channel velocity profiles measured at the free surface indicate that the range of the ``wall effect" to extends to a greater distance from each wall with increasing gap width, even when the lengths are scaled by the gap width. This, however, may be a reflection of the fact the flow rate varies significantly across the gap. \citet{cou05} measured the instantaneous velocity profiles during an avalanche in a rotating cylinder. The profiles obtained at different times at the centre of the cylinder are self-similar and have an exponential dependence on the depth. The linear part of the profile, found for steady flows, is not seen. \citet{jain02} have reported measurements of the granular temperature, which is an important input in some models for the flow.  The qualitative features of the flow remain unchanged even when the particles are completely immersed in a liquid \citep*{jain04}. 

The simplest models for surface flow comprise a depth averaged mass balance equation in the flowing layer with a phenomenological equation for the flux from the bed to the layer assumed to be determined by the local angle of the surface \citep{bouc94,meh94,bout96,bout98a}. More detailed depth averaged continuum models also yield similar equations \citep{kha97a,dou99,kha01b} and agree well with experimental results \citep{kha01b,kha01c}. The main difference in these models is the stress constitutive equation used. One approach is to assume the shear rate to be a known constant which eliminates the need for the stress constitutive equation \citep{dou99,mak99b}. \citet{raj03} gave a physical justification for a constant shear rate based on an assumption that a single collision entirely dissipates the momentum of both colliding particles. This yields an estimate for the shear rate as $\dot{\gamma}\sim(g\sin\beta/d)^{1/2}$, where $\beta$ is the local surface angle of flow, $d$ is the particle diameter and $g$ is acceleration due to gravity. Such an approach gives good predictions of experimental results for rotating cylinder flow at low rotational speeds \citep{mak99b}. A phenomenological equation for the stress based on frictional and collisional contributions \citep{kha97a,orp01,kha01b} gives a somewhat different expression for the shear rate
\begin{equation}  \label{eq:shear}
\dot{\gamma} = \left[\frac{g\sin(\beta_m-\beta_{s})}{cd\cos\beta_{s}}\right]^{1/2}
\end{equation}
where $\beta_m$ is the angle of static friction, $\beta_{s}$ is the static angle of repose (angle of dynamic friction) and $c$ is a constant. From the analysis it can be shown that the free surface angle for a steady fully developed flow with no interchange is equal to $\beta_m$. The continuum model based on the phenomenological stress equation gives good predictions of the layer thickness profile and shear rate for both heap flows \citep{kha01b} and rotating cylinder flows \citep{orp01}. The model predicts a nearly constant shear rate at low flow rates \citep{kha01c}. 

More detailed rheological models for the dense granular flow, typical of surface flows, have also been proposed based on different approaches. \citet{sav98} developed a kinetic theory based on strain-rate fluctuations which were related to the granular temperature. The resulting governing equations contain a viscosity term that shows a decrease with an increase in the granular temperature, similar to that for a liquid. \citet{boc01} modified granular kinetic theory equations to give a sharper increase in the granular viscosity with solids volume fraction. The model gave good predictions of the velocity profile in a Couette flow. A two phase solid-liquid model was proposed by \citet{aran02}, with the fraction of solid phase present locally determined by an order parameter. This model is described in more detail in section~\ref{sec:comparison-theory} and predictions are compared to experimental results. \citet{joss04} have proposed a continuum model based on stress constitutive equations which depend on the local volume fraction. The model gives a reasonable qualitative description of surface flows. Dense granular flows have also been described by Cosserat type constitutive models in which local couple stresses due to particle rotations are taken into account. A Cosserat model was proposed by \citet{moh02} and predictions of the model for the velocity profile were shown in good agreement with measurements in a vertical channel flow and dense Couette flow. Recently Pouliquen and coworkers \citep*{midi04,jop05} proposed that the local shear stress in a granular flow may be expressed in terms of a shear rate ($\dot{\gamma}$) dependent effective friction coefficient of the following form
\begin{equation}\label{eq:pouliq}
\mu=\mu_s+\frac{\mu_2-\mu_s}{I_0/I+1},
\end{equation}  
where $I=d\dot{\gamma}/(\rho_p/P)^{1/2}$ is the dimensionless shear rate, $\rho_p$ the particle density and $P$ the granular pressure. $\mu_s$, $\mu_2$ and $I_0$ are model parameters. The model is found to work well for flow on a rough inclined plane and heap flow \citep*{jop05}. We compare our experimental reults to this model as well. 

All the models discussed above are based on a \textit{local} stress constitutive equation. The high density of the flow results in multi-particle contacts and thus there is a distinct possibility of non-local transfer of stresses. Several theories have explored such ideas. \citet*{poul01} considered the stress transfer by fluctuations in the bed.  The fluctuations decay with distance from their point of origin. The model
gives good predictions of data for dense flows in verical and inclined channels. \citet*{mills99} and \citet{bon03} have considered dense shear flows in which near vertical particle chains are formed due to gravity forces. The model gives good predictions of measured velocity profiles. Predictions of this model are also compared to our experimental data below. 

The rheology of surface granular flows has not been experimentally explored in detail as yet and is the focus of this work. The objective of the present work is to carry out an experimental study and analysis of the flow and rheology of a steady, fully-developed surface granular flow. A quasi-2-dimensional rotating cylinder system (gap width 5-10 particle diameters) is used and the measurements are made in the region near the cylinder axis where the flow is fully developed and steady.  Experiments are carried out over a range of particle sizes and rotational speeds, and the data are analyzed by scaling, and comparison to existing models. Section~\ref{sec:experimental-details} gives the details of the experimental system and methods used. The results are discussed in section~\ref{sec:results-discussion} followed by conclusions of the work in section~\ref{sec:conclusions}.

\section[Expdetails]{Experimental details}
\label{sec:experimental-details}

We first describe the experimental system used for this work. The experimental method based on digital photography is discussed next, followed by a description of the technique used for analyzing the digital images.

\subsection[system]{System}
\label{sec:system-used}

Experiments are carried out in quasi-2-dimensional aluminium cylinders (length $b=1$ or 2 cm) of radius 16 cm. The end walls are made of glass for visualizing the particle motion. A computer controlled stepper motor with a sufficiently small step is used to rotate the cylinders.  Stainless steel (SS) balls of three different sizes
with diameters ($d$) 1, 2, and 3 mm are used in the experiments. A few experiments are also carried out using $2$ mm brass balls to investigate the effect of material characteristics. Brass is a relatively heavier and softer material as compared to SS. The properties of all the particles used are listed in table~\ref{tab:part-prop}.  All the particles used in the experiments are highly spherical in shape (the SS balls are used for manufacturing
ball bearings). The surface of all the particles is smooth and reflective which facilitates the image analysis method used in this work.

\begin{table}
\centering
\begin{tabular}{cccccc} 
Material & Diameter ($d$) & Mass ($m$) & Particle&Bulk  & Young's  \\
 & & &density ($\rho_{p}$)  &density ($\rho_b$) &Modulus\\
         & mm & mg & kg/m$^3$ & kg/m$^3$& GPa \\ \hline
         & 1 & 4.2 $\pm$ 0.06 & &4973& \\
SS    & 2 & 33.5 $\pm$ 0.18 & 8000 &4899 &200 \\
         & 3 & 112.5 $\pm$ 0.18 & &4725 &\\
Brass & 2 &35.0 $\pm$ 0.25 & 8500 &5205 &120
\end{tabular}
\caption{Geometrical and physical properties of particles.} \label{tab:part-prop}
\end{table}

\begin{figure}
\centering {\includegraphics[width=2in]{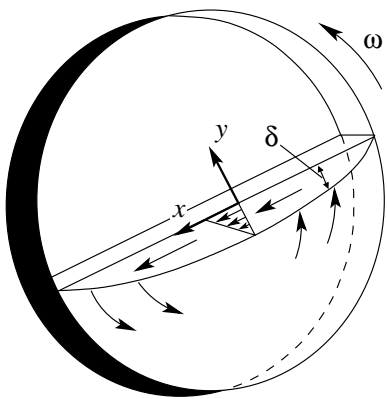}}
\caption{Schematic view of the flow in a rotating cylinder. comprising a surface flowing layer of thickness $\delta(x)$ and a packed ped of particles rotating with the cylinder. The coordinate system employed in the analysis is shown.}  \label{fig:schem}
\end{figure}

Experiments are carried out at different rotational speeds ($\omega = 2-9$ rpm) such that the flow is always in the rolling regime. The typical flow pattern, comprising a surface flowing layer of varying thickness ($\delta$($x$)) and the bed rotating at the angular speed of the drum ($\omega$), is shown schematically in figure~\ref{fig:schem}. The particles are heavy enough and conductive so that static charge effects are negligible.  All the experiments are carried out with the cylinder half-full of particles (i.e., a fill fraction of 50\%).

\subsection[Imageacq]{Experimental method}
\label{sec:image-acquisition}

\begin{figure}
\centering {\includegraphics[width=2in]{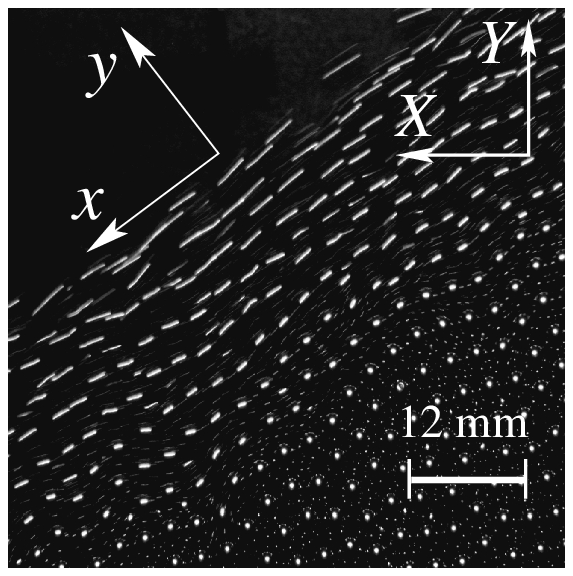}}
\caption{Flow of SS balls of diameter $2$ mm for an angular speed ($\omega$) of 3 rpm. Streaks correspond to the particle displacement during the time the shutter is open $\Delta t = 1/500$ s. $x$,$y$ correspond to the coordinates with the $y$-axis along the free surface of the flowing layer, while $X$,$Y$ correspond to the coordinates in the laboratory reference frame.} \label{fig:streak}
\end{figure}

The measurements in the flowing layer are made near the center of the cylinder ($x=0$, in figure~\ref{fig:schem}), where the layer thickness is maximum and the flow is nearly unidirectional and non-accelerating (velocity is nearly invariant in the $x$-direction). The images are taken keeping the camera close to the face plate of the cylinder and framing a small region of flow. The size of the recorded region is $2560 \times 1920$ pixels, with one pixel corresponding to $0.016-0.03$ mm depending on the distance of the camera from the cylinder. A point source of light is directed nearly parallel to the face plate of the cylinder so as to illuminate only the front layer of the flowing particles. Owing to the high sphericity and the polished surface of the particles, a well-defined reflection is produced from each stationary particle (see figure~\ref{fig:streak}, lower right corner).  Each moving particle thus generates a streak of definite length depending on its speed and the shutter speed of the camera. Figure~\ref{fig:streak} shows a typical image with streaks of different lengths in the flowing layer.  The velocity of each particle is determined from the length and the orientation of the streak formed; the analysis method is explained in \ref{sec:image-analysis}.  This method of velocity determination is similar to that used by \citet{raj95}.

Images are taken for a range of camera shutter speeds ($1/15-1/1000$ s) so as to account for the varying velocity across the flowing layer. About one hundred images are taken for each shutter speed with an overall of thousand images combined over different shutter speeds.  Each experiment is carried out for about 500 revolutions with two images taken per revolution.  Thus the flow is sampled over a considerable time period ($\sim$ 3 h). The images are analyzed to determine the velocity and the number density profiles across the layer.

\begin{figure}
\centering {\includegraphics[width=5.0in]{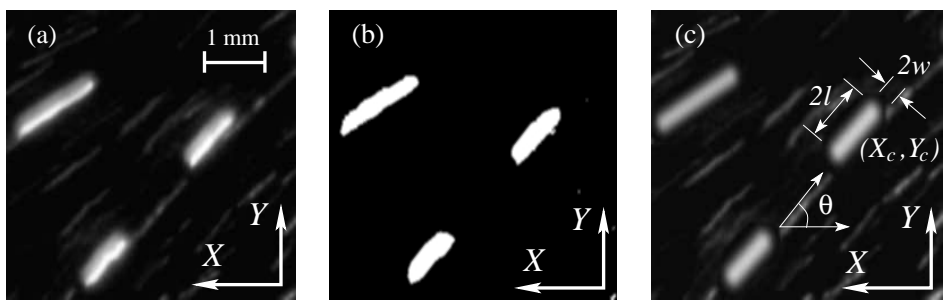}}
\caption{Image analysis of streaks. (a) Original image of the streaks. (b) Thresholded streaks with pixels identified in each streak. (c) Fitting of the function (\ref{eq:intfunc}) to one streak.  $X,Y$: coordinate axis in the laboratory reference frame. $X_{c},Y_c$: Position of the streak in this frame of reference.  $2l$ and $2w$ are the length and width of the streak, respectively.  $\theta$ is the orientation of the streak with the horizontal.} \label{fig:streakanal}
\end{figure}

\subsection[Imganal]{Image analysis}
\label{sec:image-analysis}

The image analysis technique involves identification of the streaks in the front layer followed by determination of the length and the orientation of each identified streak. Computer codes have been written to manipulate and analyze the digital images. Details are given below.

Each image is first converted to black and white by putting intensity values of pixels greater than a threshold to white and the rest to black. The threshold is chosen so that the streaks from the front layer are converted to white and the streaks in the back layers, which are less illuminated, to black. All the photographs are taken with a black background so as to make the thresholding easier. The thresholded image is scanned row by row to identify clusters of contiguous white pixels. Clusters smaller than a specified size ($75$ pixels) are due to reflection of the light from the periphery of the particle surface and are eliminated. Clusters larger than this size are taken to be streaks. Figure~\ref{fig:streakanal}a shows three magnified streaks in the front layer and a number of faint streaks from the back layers. Figure~\ref{fig:streakanal}b shows the thresholded image with faint streaks eliminated.

In the next stage of the analysis, the position, length and angle of each streak is determined by fitting a parameterized intensity function to the intensity values of the streak pixels (and an immediate neighbourhood) in the original image. An intensity function of the following form is used
\begin{subeqnarray}\label{eq:intfunc}
I_{th} &=& C e^{-\bar{y}^{2}/4w^{2}}\:\:\:\:\:\:\:\:\:\:\:\:\:\:\:\:\:\:\:\:|\bar{x}| \leq l \\
        &=& C e^{[-\bar{y}^{2}-(\bar{x}-l)^{2}]/4w^{2}} \:\:\:\:\: |\bar{x}|> l
\end{subeqnarray}
where $\bar{x} = (X-X_{c})\cos\theta + (Y-\lambda)\sin\theta$, $\bar{y} = (Y-Y_c)\cos\theta - (X-X_{c})\sin\theta$, $2l$ is the nominal length of the streak, $2w$ is the nominal width of the streak, $\theta$ is the orientation of the streak with the horizontal and $C$ is the maximum intensity value obtained at the centroid of the streak.  ($X_{c},Y_c$) is the position of the centroid of the streak in laboratory coordinates (Fig.~\ref{fig:streakanal}c).

Initial estimates of the position ($X_{c},Y_c$), length ($2l$), width ($2w$) and orientation angle ($\theta$) are first obtained for an identified streak from the thresholded image. The values are refined by minimizing the following function for each streak using Powell's method \citep{num94}
\begin{equation}\label{eq:func}
f = \sum(I_{th} - I_{image})^{2},
\end{equation}
where $I_{image}$ is the intensity value for a pixel in the original image at that point (figure~\ref{fig:streakanal}a).  The summation in (\ref{eq:func}) is carried over all the pixels of the streak together with six layers of pixels added to the cluster. Figure~\ref{fig:streakanal}c shows the computed intensity values plotted on the original image using the fitted parameters.

The velocity magnitude is $v=2L/\Delta t$, where $\Delta t$ is the time interval corresponding to the shutter speed set in the camera, and the velocity components are calculated from the orientation angle, $\theta$.  $L=L(l,w)$ is the distance moved by the particle in the time interval $\Delta t$ and is obtained as a function of the streak length and width using calibration experiments described below. The position corresponding to this velocity vector is taken to be the centroid of the streak ($X_{c},Y_c$).  A similar procedure is applied for every streak identified in an image.  Only streaks with lengths in the range $20-70$ pixels are considered for analysis. This is because streaks longer than $70$ pixels tend to overlap with other streaks, while those shorter than $20$ pixels result in an inaccurate determination of the orientation angle ($\theta$). The use of different shutter speeds gives streaks with lengths within this range at each depth in the layer. Streaks which are cut off at the edges of an image are not considered for velocity calculation.

The analysis results from all the images are combined and the coordinate positions for all the selected streaks are rotated by angle $\theta_{avg}$.  This angle is determined by averaging the angle of orientation ($\theta$) for all the streaks in the selected region of the flowing layer, and represents the orientation of the flowing layer with the horizontal. The components of the instantaneous velocity parallel ($c_{x}$) and perpendicular ($c_{y}$) to the flow direction are determined for every streak from its orientation angle $\theta$. The region shown in figure~\ref{fig:streak} is then divided into bins of depth equal to the particle diameter and length $20$ mm parallel to the flowing layer. The profiles of the averaged parallel ($v_{x}$) and perpendicular ($v_{y}$) components of velocity are then generated by averaging the instantaneous velocities ($c_{x},c_{y}$) over all streaks in each bin.

In a typical measured velocity profile, the average $x$-velocity ($v_{x}$) decreases and then starts to increase with depth in the flowing layer from the free surface. This is because velocity decreases and eventually becomes negative (due to bed rotation) with decreasing $y$. However, our technique cannot give the direction of the velocity vector. Hence we assume all velocities below the minimum velocity point to be negative. 

The root mean square (r.m.s.) velocity and its components in $x$ and $y$ direction are calculated from the instantaneous velocities using following equations
\begin{subeqnarray}
  \label{eq:rmsvel}
  u_{x} &=& \sqrt{\overline{c_{x}^{2}} - (\overline{c_{x}})^{2}} \\
  u_{y} &=& \sqrt{\overline{c_{y}^{2}} - (\overline{c_{y}})^{2}} \\
  u &=& \sqrt{u_{x}^{2} + u_{y}^{2}}
\end{subeqnarray}
where ($\overline{\oplus}$) represents an average over all streaks in a bin corresponding to a $y$-coordinate.  The number density profiles are obtained from a separate set of images at a high shutter speed ($1/2000$ s). Images taken at such high shutter speeds ensure that all streaks are short ($\leq 30$ pixels) and do not overlap so that all streaks in an image can be used for analysis. This is required to get an accurate number count. Images are analyzed in the same manner as mentioned above, but only the centroid positions are noted. The number density for any $y$ is determined by counting the number of particles in each bin. A source of inaccuracy is that particles from back layers may also be counted which leads to an overestimate of the number densities. This is particularly the case for the low number density region near the free surface.  Further, the measurements represent the densities near the wall which are different from those in the bulk. 

\begin{figure}
\centering {\includegraphics[width=3.0in]{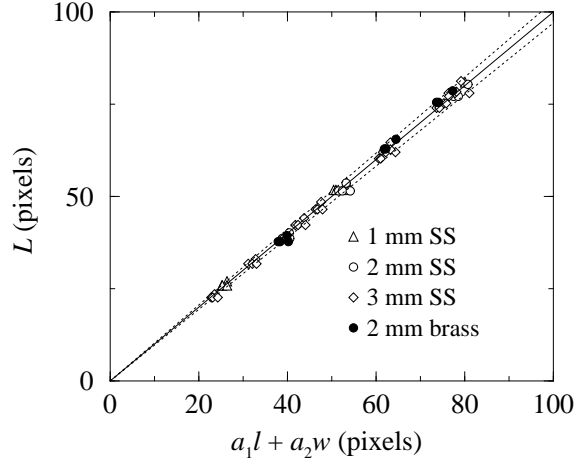}}
\caption{Calibration results for determining streak lengths. Variation of actual streak length ($L$) with a linear sum of length ($l$) and width ($w$) of a streak measured by image analysis. $a_{1} = 2.05$ and $a_{2} = 1.86$ are the constants obtained by linear regression of the experimental data. Dotted lines represent the $3\%$ deviation in calculation of $L$.}   \label{fig:calib}
\end{figure}

Experiments are carried out to calibrate the analysis technique. A single particle is glued to the face plate of the cylinder from the inner side at a position close to the periphery of the cylinder.  The radial position of the particle is accurately determined from which the velocity of the particle can be obtained since the angular speed of the cylinder is known.  The cylinder is rotated at different angular speeds and images of this particle are taken at different camera shutter speeds ($1/30$ s - $1/500$ s). The size of the recorded region and the lighting conditions are the same as those used for the measurements in the flowing layer. Further, the rotational speed of the cylinder and the shutter speeds are adjusted so as to generate streaks of three different lengths. The maximum and the minimum lengths of the streaks span the entire range of streak lengths considered for velocity measurements in the flowing layer. The experiments are carried out for all the particles used in the experiments. The length ($2l$), width ($2w$), position ($X_{c},Y_c$) and orientation ($\theta$) of this single streak are determined in each case using the same analysis procedure as described earlier. The actual distance traveled by the particle at radial distance $r$ ($L = \omega r \Delta t$) is known from the linear velocity of the particle ($\omega r$) and the shutter speed used ($\Delta t$). A linear relation of the following form is assumed to hold between the actual length ($L$) and measured length ($l$) and width ($w$) of each streak
\begin{equation}
  \label{eq:calib}
  L = a_{1} l + a_{2} w.
\end{equation}
The constants $a_{1}$ and $a_{2}$ are determined by a least-squares fit to the experimental data combined for particles of different sizes and materials.  The fit yields $a_{1} = 2.05$ and $a_{2} = 1.86$ with the standard error of about 1.1 pixels.  Figure~\ref{fig:calib} shows a comparison of the measured and fitted streaks. The experimental data are well correlated by (\ref{eq:calib}) and indicate that the actual distance traveled by a particle depends on both the length and width parameters of the intensity function.  The error in calculating $L$ is less than $3\%$ (figure~\ref{fig:calib}) and represents the typical error in measurements of velocities in the flowing layer.

\begin{figure}
\centering {\includegraphics[width=3.0in]{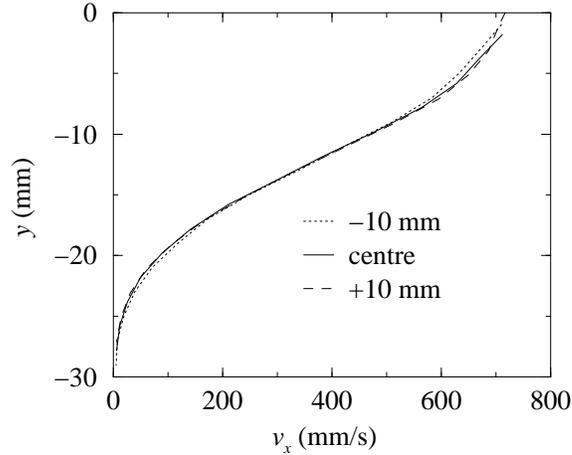}}
\caption{Mean velocity profiles measured at three locations along the layer as indicated in the legend, for 2 mm particles with the cylinder rotated at 3 rpm.}   \label{fig:vxpm}
\end{figure}

The dynamic angle of repose ($\beta_{m}$) is taken to be the angle by which the data is rotated ($\theta_{avg}$) to obtain $v_y=0$. The static angle of repose ($\beta_{s}$) is taken to be the angle of the stationary free surface when the rotation of the cylinder, operating under steady flow conditions, is stopped. The values of $\beta_{m}$ and $\beta_{s}$ (averaged over six images) are reported in table~\ref{tab:betam-betas} for all the cases studied. We also checked that the flow was fully developed by measuring the velocity profile at two additional locations: one 10 mm upstream and one 10 mm downstream of the layer midpoint ($x=0$). Figure \ref{fig:vxpm} shows the profiles of the mean velocity at the three locations for one case (2 mm particles in the cylinder rotated at 3 rpm). The profiles are identical within experimental error, verifying the the flow is fully-developed. 

\begin{table}
\centering
\begin{tabular}{cccccc}
Material & $d$ & $\omega$ & $\beta_{m}$ &$\beta_{s}$ &$q$ \\ 
    & (mm) & (rpm) & (deg.) & (deg.) &(cm$^2$/s)\\ \hline
    & $1$ & $2$ & 25.6 & 16.0 &26.8\\
    & $1$ & $3$ & 26.6 & 16.0 &40.2\\
    SS & $2$ & $3$ & 29.1 & 20.1 &\\
    & $2$ & $6$ & 35.7 & 20.1 &80.4\\
   & $2$ & $9$ & 42.8 & 20.1 &120.6\\
    & $3$ & $3$ & 26.3 & 18.4 &\\
    Brass & $2$ & $3$ & 27.8 & 18.3 & 
\end{tabular}
\caption{Experimentally determined values of the angles $\beta_{m}$ and $\beta_{s}$, and calculated volumetric flow rate per unit width ($q$) for the different cases studied.}\label{tab:betam-betas}
\end{table}

\section[Results]{Results and discussion}
\label{sec:results-discussion}

We first present the experimental results obtained for varying system parameters. Scaling and the analysis of the experimental profiles are presented next. The rheology of the flow is discussed in \ref{sec:rheol}. Finally, we compare our experimental velocity profiles with predictions of the models of \citet{bon03}, \citet{aran02} and Pouliquen and coworkers \citep*{jop05,midi04}. 

\subsection{Base data}
\label{sec:base-data}

The primary parameters varied in the experiments are the cylinder rotational speed ($\omega$) and the particle diameter ($d$), and the results presented below are in terms of these parameters. There is, however, a direct relation between the cylinder rotational speed and the local volumetric flow rate per unit gap width ($q$). A mass balance for a half-full cylinder, assuming that the bulk density in the bed and the layer are the same, yields $q=\omega R^2/2$, and calculated values are given in table~\ref{tab:betam-betas}. Increasing the rotational speed is thus equivalent to increasing the local flow rate.
 
\begin{figure}
\centering {\includegraphics[width=5.3in]{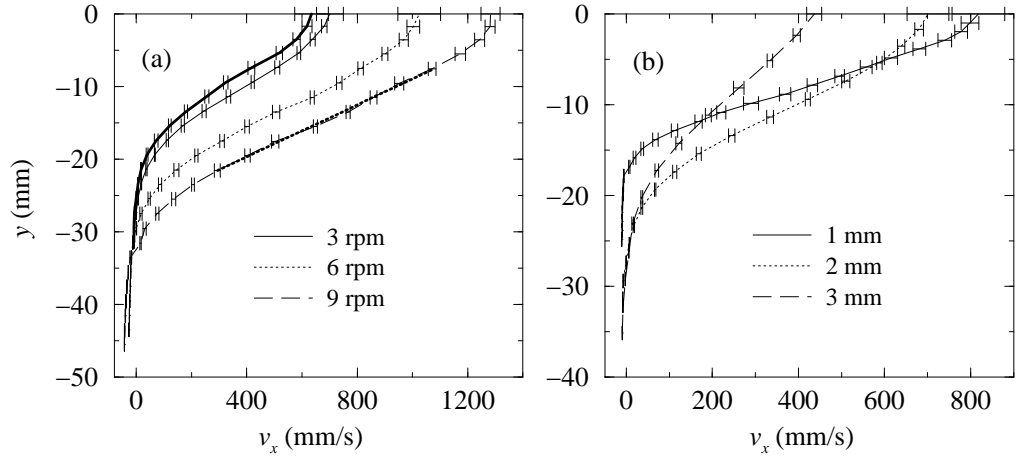}}
\caption{Variation of $x$-component of mean velocity ($v_{x}$) with depth ($y$) in the layer. (a) Thick solid line denote experimental data for $2$ mm brass particles rotated at 3 rpm. Other lines denote data for $2$ mm SS balls at different rotational speeds. Thick dotted line shows a straight line fitted to the linear portion of the 9 rpm profile. (b) Thick solid line denote data for $1$ mm SS balls rotated at $2$ rpm. Other lines denote data for SS balls rotated at 3 rpm. Error bars denote standard deviation over 10 sets.}  \label{fig:vxvsy}
\end{figure}

\begin{figure}
\centering {\includegraphics[width=3in]{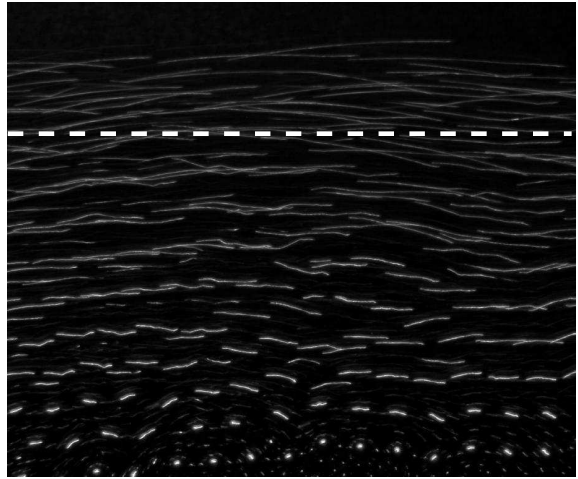}}
\caption{Typical time lapse image showing ballistic trajectories of particles in the low density region (above the dashed line).}  \label{fig:saltating}
\end{figure}

The variation of the $x$ component of the mean velocity ($v_{x}$) with depth ($y$) in the layer is shown in figure~\ref{fig:vxvsy} for different rotational speeds, particle sizes and materials studied. The error bars denote the standard deviation over ten different data sets. All the measured quantities are plotted from $y=0$ at the free surface of the flow down to about 3-4 particle diameters below the point corresponding to the minimum absolute velocity. The free surface is taken to be the point at which the measured number density is close to zero ($ n < 10^{-3}$ mm$^{-2}$). The profiles are linear over most of the layer depth for all the cases studied, with an exponential decay near the base of the flowing layer. The negative velocities correspond to the particles in the rotating bed which move in a direction opposite to that of the flowing layer. Further, the profiles tend to flatten out near the free surface. This is a low density region comprising saltating particles. The ballistic trajectories in this region are clearly seen in figure~\ref{fig:saltating}, which is an image of the flow taken at a low shutter speed.

The magnitude of the maximum velocity and the depth of the layer increase with the rotational speed for a fixed particle size (figure~\ref{fig:vxvsy}a). The length of central linear region of the velocity profile increases with increasing rotational speed, while the length of the exponential region near the bottom of the layer is nearly constant for all the cases. The shear rate ($\dot{\gamma} = dv_{x}/dy$) corresponding to the linear portion increases with rotational speed. The profiles for SS and brass particles are close to each other indicating that particle material parameters like coefficient of restitution and particle roughness have a minor effect on the flow. This is in agreement with our earlier work \citep{orp01} and that of \citet{jain02} in rotating cylinders. The maximum velocity decreases and the depth of flowing layer increases with increasing particle diameter (figure~\ref{fig:vxvsy}b). The shear rate, thus, decreases with increasing particle size. 

Qualitatively similar velocity profiles have been reported in the literature for different experimental systems. A linear and an exponential region of velocity profile was observed in rotating cylinders for various rotational speeds studied \citep{bon02a,longo02,zan02}. The trends obtained in figure~\ref{fig:vxvsy}b for varying particle size are similar to those reported by \citet{jain02}. The system size used in all these cases is comparable to our experimental system. The velocity profile in the lower half of the layer resembles the profiles obtained by \citet{mueth00} and \citet{boc01} for Couette flows.  In some of the studies mentioned above \citep{raj95,bon02a}, the flattened upper low density region was not obtained. 

\begin{figure}
\centering {\includegraphics[width=5.3in]{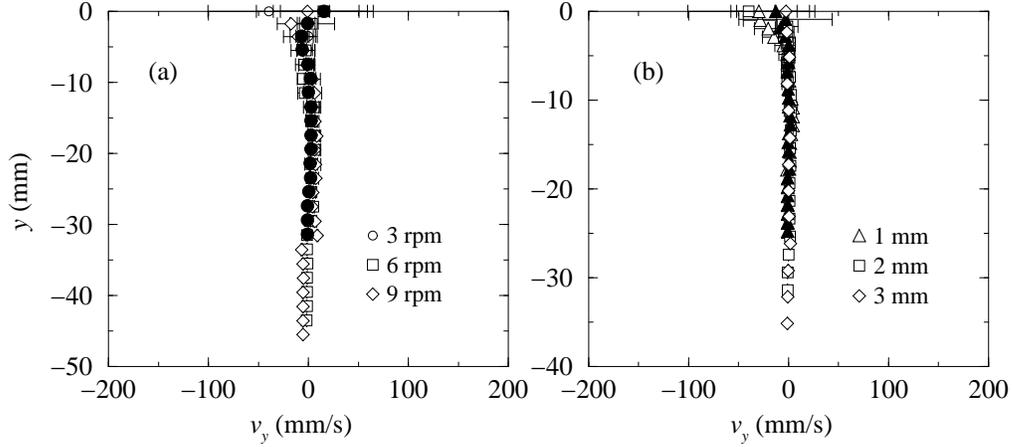}}
\caption{Variation of $y$-component of mean velocity ($v_{y}$) with depth ($y$) in the flowing layer. (a) Open symbols denote experimental data for $2$ mm SS balls. Filled symbols denote data for $2$ mm brass particles rotated at 3 rpm. (b) Open symbols denote experimental data for SS balls for a rotational speed of 3 rpm. Filled  symbols denote data for $1$ mm SS balls rotated at $2$ rpm. Error bars denote standard deviation over 10 sets.} \label{fig:vyvsy}
\end{figure}

\begin{figure}
\centering {\includegraphics[width=5.3in]{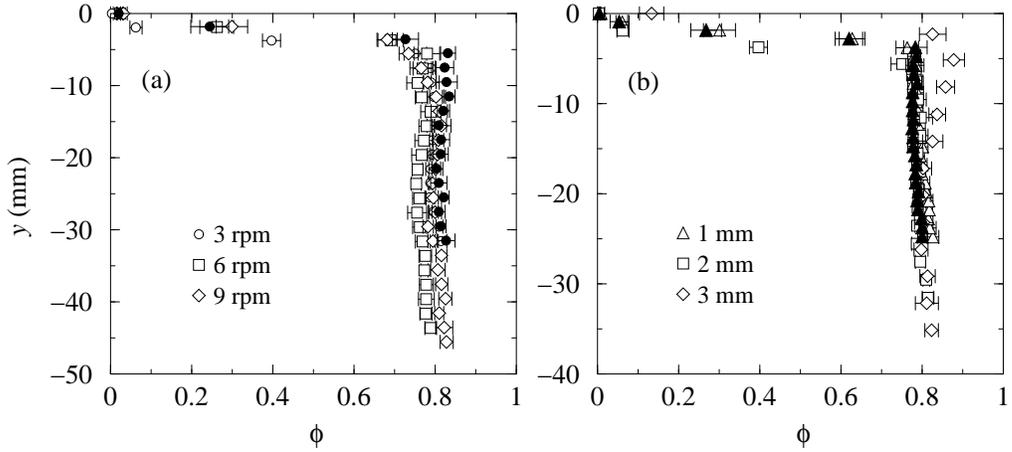}}
\caption{Variation of area fraction ($\phi$) with depth ($y$) in the flowing layer. (a) Open symbols denote experimental data for $2$ mm SS balls. Filled symbols denote data for $2$ mm brass particles rotated at 3 rpm. (b) Open symbols denote experimental data for SS balls for a rotational speed of 3 rpm. Filled symbols denote data for $1$ mm SS balls rotated at $2$ rpm. Error bars denote standard deviation over 10 sets.} \label{fig:frac}
\end{figure}

The magnitude of the $y$-component of the mean velocity ($v_{y}$) is close to zero throughout the depth of the layer for all the cases (figure~\ref{fig:vyvsy}). This confirms that the flow is nearly unidirectional (invariant in the $x$ direction) in the region of measurement. The profiles of the  projected area fraction of the particles $\phi = n \pi d^{2}/4 A$, where $n$ is the number of particles counted per bin and $A$ is the area of the bin are shown in figure~\ref{fig:frac}. The area fraction is nearly constant over the entire layer for all the particles with a rapid drop near the free surface. No appreciable change in the magnitude is observed for different rotational speeds and materials. The area fractions are significantly smaller than the hexagonal close packed limit ($\phi_c = 0.866$). Nearly constant surface densities across the flowing layer were also observed by \citet{raj95}, \citet{bon01} and \citet{jain02}.

\begin{figure}
\centering {\includegraphics[width=5.3in]{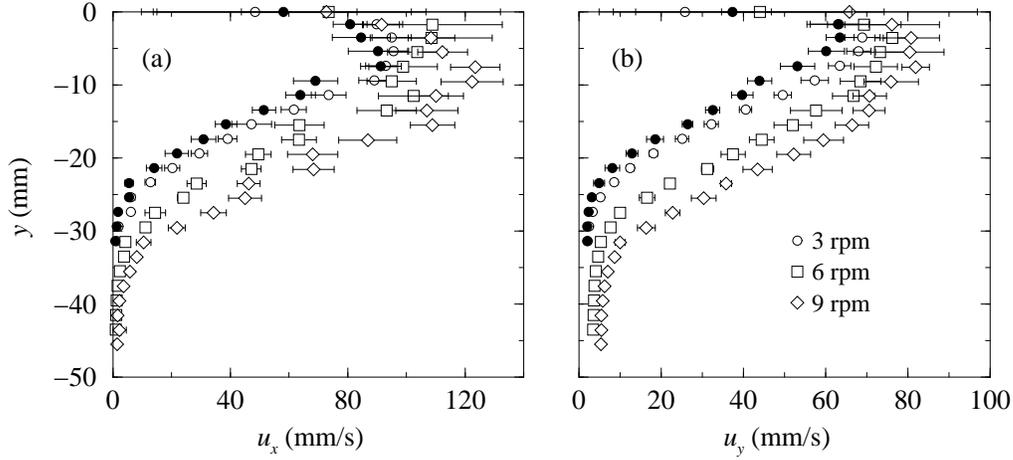}}
\caption{Variation of the components of r.m.s.\ velocity with depth in the flowing layer. Open symbols denote experimental data for $2$ mm SS balls. Filled symbols denote experimental data for $2$ mm brass particles rotated at 3 rpm. Error bars denote standard deviation over 10 sets.} \label{fig:rpmtxty}
\end{figure}

\begin{figure}
\centering {\includegraphics[width=5.3in]{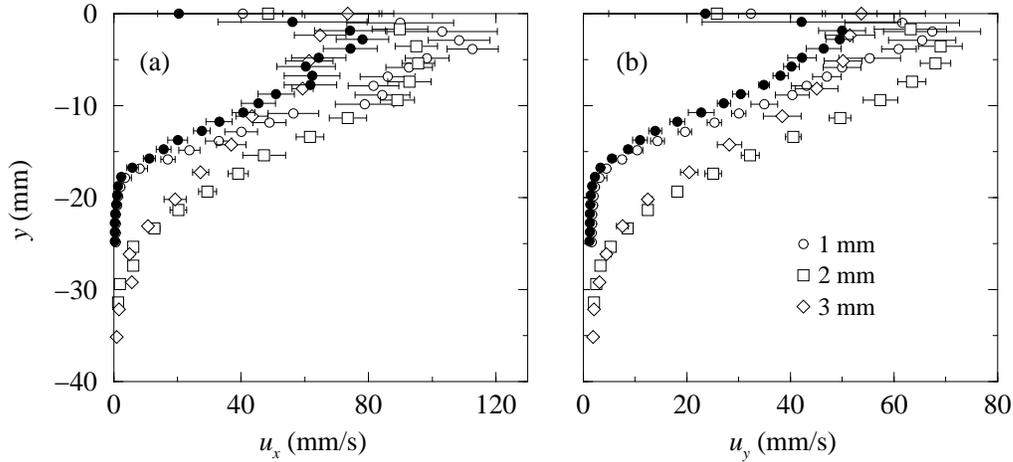}}
\caption{Variation of the components of r.m.s.\ velocity with depth in the flowing layer. Open symbols denote data for SS balls rotated at $3$ rpm. Filled symbols denote data for $1$ mm SS balls rotated at $2$ rpm. Error bars denote standard deviation over 10 sets.} \label{fig:sizetxty}
\end{figure}

\begin{figure}
\centering {\includegraphics[width=5.3in]{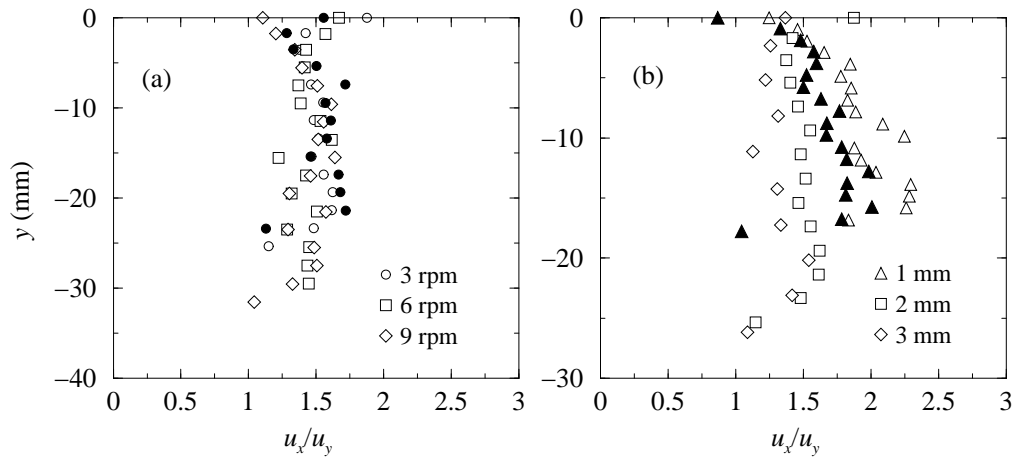}}
\caption{Variation of the ratio of the r.m.s.\ velocity components across the flowing layer. (a) Open symbols denote data for $2$ mm SS balls. Filled symbols denote data for $2$ mm brass particles. (b) Open symbols denote data for SS balls rotated at $3$ rpm. Filled symbols denote data for $1$ mm SS balls rotated at $2$ rpm.} \label{fig:txtyratio}
\end{figure}

The components of the r.m.s.\ velocities ($u_{x}$,$u_{y}$) across the flowing layer are shown in figures~\ref{fig:rpmtxty} and \ref{fig:sizetxty}. In both cases, the r.m.s.\ velocity profiles are nearly linear over a large portion of the layer with a flattening near the free surface and an exponential decay near the base of the flowing layer. The sizes of the flattened region as well as the region of exponential decay are larger for the r.m.s.\ velocity profiles than the corresponding regions for the mean velocity profiles. The r.m.s.\ velocities are smaller by nearly an order of magnitude as compared to the mean velocities. The relative error is thus larger in this case as compared to the case of mean velocities, as indicated by the error bars. The r.m.s.\ velocities increase with rotational speed for a fixed particle size. The slope of the linear region in each case is nearly independent of the rotational speed, but increases with particle diameter (i.e., $du_{x}/dy, du_{y}/dy$ decrease). The r.m.s.\ velocities for the brass balls are quite close to those for steel which again indicates that the flow behaviour is nearly independent of the material properties. The anisotropy in the r.m.s.\ velocities is clearly evident from the graphs and the r.m.s.\ velocity component in the flow direction is larger by about $50\%$ than that normal to the flow direction.  A linear and a flattened region in the r.m.s.\ velocity profiles was also obtained by \citet{jain02} in their experiments in rotating cylinders but they found the r.m.s.\ velocities to be about $1/3$rd of the mean velocity near the free surface. These values are significantly larger than the r.m.s.\ velocities obtained here. 

The degree of anisotropy in the r.m.s.\ velocities across the flowing layer for all the cases studied is shown in figure~\ref{fig:txtyratio}. The ratio $u_{x}/u_{y}$ ($\approx 1.5$) is nearly constant across the flowing layer and independent of rotational speed, while it decreases slightly with an increase in the particle diameter. The data for brass and SS balls is nearly the same. \citet*{azanza99} found the ratio of the r.m.s.\ velocity components to be $1.45$ for 2-dimensional chute flows, while a ratio of $1.3$ was obtained by \citet{boc01} for Couette flows. Anisotropy in the r.m.s.\ velocity components was also observed by \citet{jain02} for the experiments in rotating cylinders. 

We obtain the stress profiles in the flowing layer from a stress balance taking into account the contribution of wall friction. Assuming the flow to be steady and fully developed, the Cauchy equations averaged across the gap width ($b$) are 
\begin{subeqnarray}
0&=&\frac{d\left\langle\tau_{yx}\right\rangle}{dy}+\frac{1}{b}\left.\tau_{zx}\right|_0^b
+\left\langle\rho\right\rangle g\sin\beta_m\\
0&=&\frac{d\left\langle\tau_{yy}\right\rangle}{dy}-\left\langle\rho\right\rangle g\cos\beta_m,
\end{subeqnarray} 
where $\langle\cdot\rangle=(1/b)\int_0^b\cdot dz$ denotes an average across the gap. The stress at the side walls due to friction is
\begin{equation}
\tau_{zx}|_0=-\tau_{zx}|_b=\tau_{zz}\mu_w,
\end{equation}  
where $\mu_w$ is the coefficient of wall friction. The coefficient of wall friction is assumed be a constant, which may overestimate the actual value. Assuming further that the variation of stresses across the gap is small, that the normal stresses are isotropic ($\tau_{zz}=\tau_{yy}$) and that the stresses vanish at the free surface we obtain
\begin{subeqnarray}\label{eq:tauyy}
\tau_{yy}&=&\rho_b\cos\beta_m\int_{-y}^0(\phi/\phi_b)dy\\
\tau_{yx}&=&\rho_b\sin\beta_m\int_{-y}^0(\phi/\phi_b)dy - 2\mu_w/b\int_{-y}^0 \tau_{yy}dy.
\end{subeqnarray} 
In the above equations $\phi_{b}$ is taken to be the constant value near the base of the flowing layer (figure~\ref{fig:frac}). The bulk density in the rotating bed is obtained as $\rho_{b} = 1/2(m_{b}/\pi R^{2}b)$,   from the measured mass of particles in the half-filled cylinder, $m_{b}$, and results are given in table~\ref{tab:part-prop}. The effective coefficient of friction for the particles at the base of the layer is obtained from (\ref{eq:tauyy}) as
\begin{equation}\label{eq:mui}
\mu_i=\frac{\tau_{yx}}{\tau_{yy}}=\tan\beta_m-\mu_w\frac{\delta_e}{b},
\end{equation}  
where $\delta_e=[2\int_{-y_s}^0\tau_{yy}dy]/[\rho_b\int_{-y_s}^0 (\phi/\phi_b)dy]$ is the effective layer thickness and $y_s$ is the position of the base of the layer, which we take to be the point at which the shear rate extrapolates to zero. Equation  (\ref{eq:mui}) is identical to one obtained by \citet{tab03} if $\delta_e$ is replaced by the measured layer thickness ($\delta$). Here we obtain $\delta_e$ by numerical intergration of the experimental profiles to avoid the ambiguity in defining the location of the free surface.  

\begin{figure}
{\centering \includegraphics[width=3in]{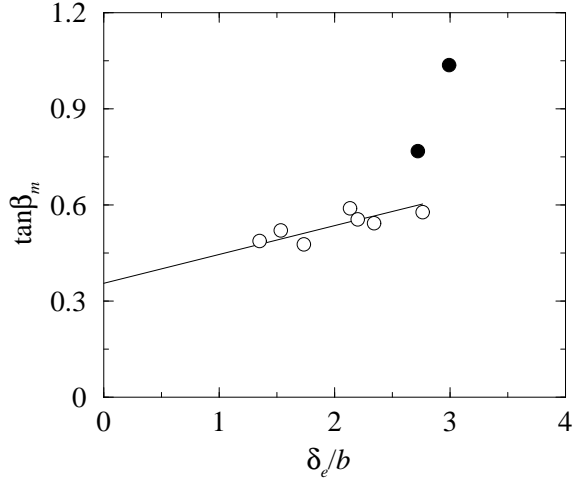} }
\caption{Variation of coefficient of static friction ($\tan\beta_m$) with scaled effective layer thickness ($\delta_e$). Open symbols: data for 3 rpm. Filled symbols: data for 6 and 9 rpm. Line is least squares fit to 3 rpm data.} \label{fig:del_betam}
\end{figure}

Equation  (\ref{eq:mui}) allows the estimation of the wall friction coefficient from experimental data. Figure~\ref{fig:del_betam} shows a plot of the measured values of $\tan\beta_m$ versus $\delta_e/b$ for all the cases studied. All the data at 3 rpm (open symols), corresponding to a fixed flow rate, fall on a straight line which yields a wall friction coefficient of $\mu_w=0.09$, which is a reasonable value for steel beads on a smooth surface. The data for higher rotational speeds (filled symbols), which correspond to significantly higher flow rates, deviate from the fitted straight line. This seems to imply that the effective coefficient of friction ($\mu_i$) may be velocity dependent, as suggested  by \citet{dou99}. In the data reported here we correct the shear stress using a wall friction coefficient of $\mu_w=0.09$ in all cases.
 
\begin{figure}
\centering {\includegraphics[width=5.3in]{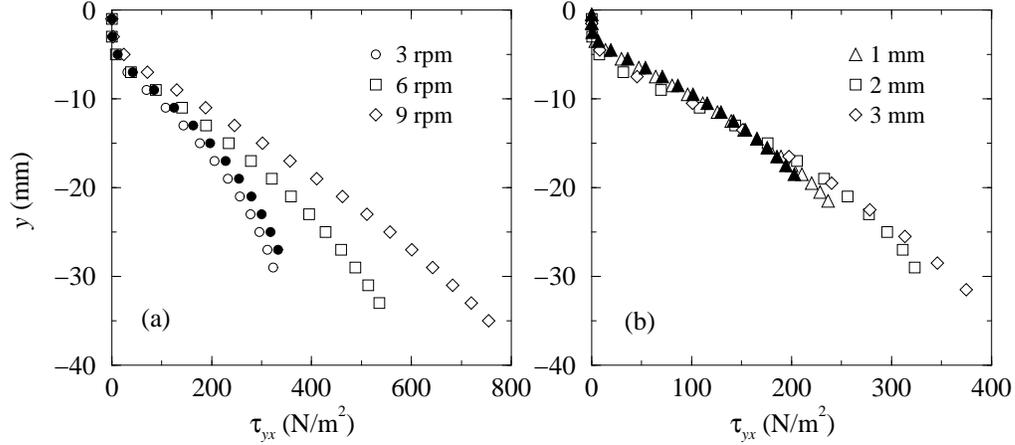}}
\caption{Shear stress ($\tau_{yx}$) profiles in the flowing layer. (a) Open symbols denote data for $2$ mm SS balls. Filled symbols denote data for $2$ mm brass particles rotated at 3 rpm. (b) Open symbols denote data for SS balls rotated at 3 rpm. Filled symbols denote data for $1$ mm SS balls rotated at 2 rpm.} \label{fig:tauyx}
\end{figure}

The profiles for the shear stress are shown in figure~\ref{fig:tauyx}. The stresses are linear across most of the layer with a slight curvature near the free surface and a deviation from linearity deeper in the layer due to wall friction.  The linear variation results from the nearly constant area fraction ($\phi$) over most of the layer.  The shear stress increases with the angular speed (figure~\ref{fig:tauyx}a), while the normal stress shows the opposite trend. This is because $\beta_m$ increases with angular speed. The shear stress profile is nearly independent of the particle size ($d$) and the material type (figure~\ref{fig:tauyx}b).

\begin{figure}
\centering {\includegraphics[width=5.3in]{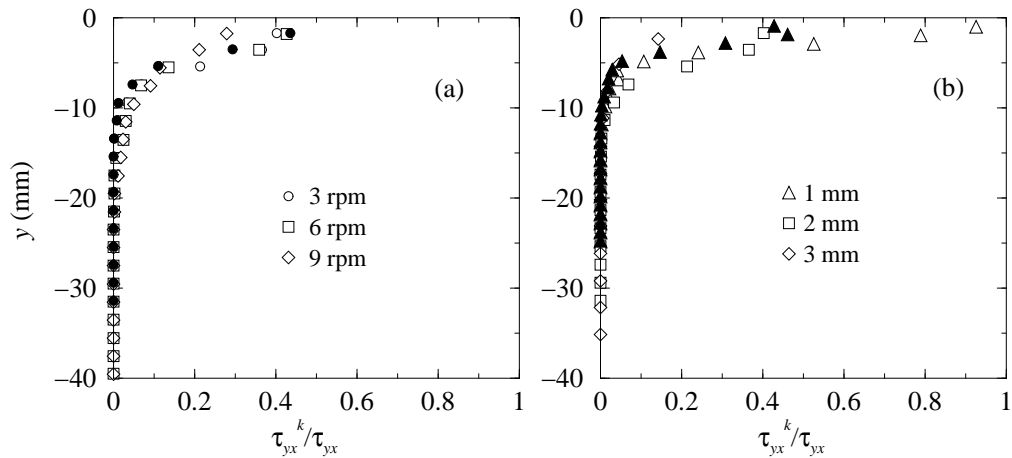}}
\caption{Variation of the ratio of streaming stress ($\tau_{yx}^{k}$) to the total shear stress ($\tau_{yx}$) with depth ($y$) in the flowing layer. (a) Open symbols denote data for $2$ mm SS balls. Filled circles denote data for $2$ mm brass particles rotated at 3 rpm. (b) Open symbols denote data for SS balls rotated at 3 rpm. Filled triangles denote data for $1$ mm SS balls rotated at 2 rpm.} \label{fig:rtauyx}
\end{figure}

The components of the streaming stress contribution to the total stress tensor are \citep{cam90} \begin{subeqnarray} \label{eq:reystress}
\tau_{yx}^{k}(y) &=&\rho(\overline{c_{x}c_{y}}-\overline{c_{x}}\;\overline{c_{y}}) \\
\tau_{xx}^{k}(y) &=&\rho(\overline{c_{x}c_{x}}-\overline{c_{x}}\;\overline{c_{x}}) \\
\tau_{yy}^{k}(y) &=&\rho(\overline{c_{y}c_{y}}-\overline{c_{y}}\;\overline{c_{y}})
\end{subeqnarray}
where $c_{x}$ and $c_{y}$ are the components of the instantaneous velocities for all the relevant particles for every $y$-coordinate and $\rho$ is the nearly constant layer density. The profiles of the streaming component of the shear stress normalized by the total shear stress are shown in figure~\ref{fig:rtauyx}. Very similar profiles are obtained for the normal stresses. Streaming stresses are a significant component of the total stress near the free surface, and are more than 20\% of the total stress in the region corresponding roughly to the flattened portion in r.m.s.\ velocity profiles. Deeper in the bed the streaming stresses become negligibly small. This behaviour is consistent with the observed particle trajectories (figure~\ref{fig:saltating}) and the density profile (figure~\ref{fig:frac}).  

\subsection{Scaling and analysis}
\label{sec:scaling-analysis}

\begin{figure}
\centering {\includegraphics[width=5.3in]{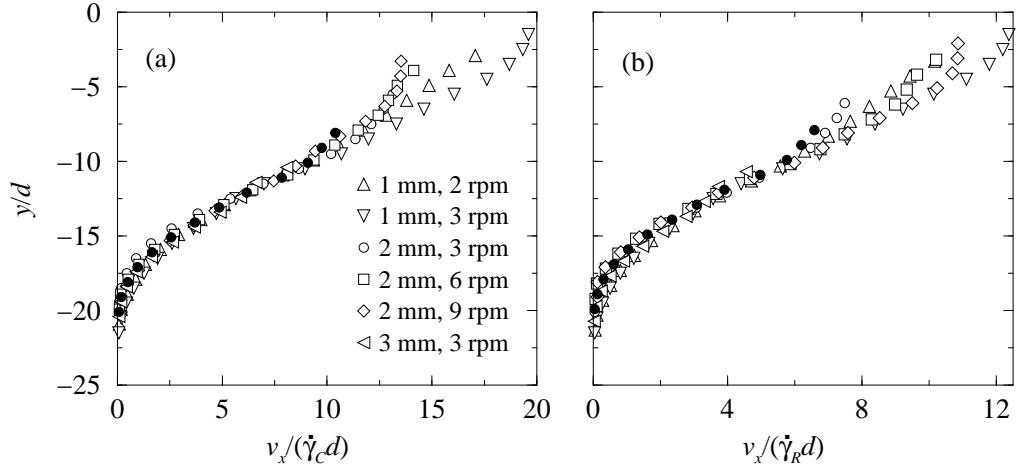}}
\caption{Variation of the rescaled mean velocity with rescaled depth in the flowing layer. Filled symbols denote data for brass particles rotated at $3$ rpm. The data for all sets are shifted after scaling so as to superimpose the profiles. (a) Scaling based on shear rate given by (\ref{eq:shear}). (b) Scaling based on shear rate of \citet{raj03}} \label{fig:vxscal}
\end{figure}

We rescale the velocity profiles using a characteristic velocity $v_{C} = \dot{\gamma_{C}}d$, with the characteristic shear rate given $\dot{\gamma_{C}} = [g\sin((\beta_m-\beta_)/d\cos\beta_s]^{1/2}$ based on (\ref{eq:shear}), and use the particle diameter ($d$) as the characteristic length. After  rescaling, the dimensionless profiles are shifted in the $y$ direction so as to superimpose the profiles. The resulting profiles are shown in figure~\ref{fig:vxscal}a. All the profiles for different rotational speeds, particle sizes and material types collapse to a single curve. The linear regions of all the profiles, in particular, superimpose very well. The scaling thus describes the variation of the mean velocity profiles with particle diameter and local flow rate quite well. Figure~\ref{fig:vxscal}b shows the same data but rescaled by the characteristic velocity based on the shear rate proposed by \citet{raj03}, $\dot{\gamma}_R=(g\sin\beta_m/d)^{1/2}$. The scaled profiles superimpose equally well in this case, after appropriately translating the profiles in the $y$ direction. From the data available it is difficult to distinguish between the two relations for the shear rate.

\begin{figure}
\centering {\includegraphics[width=5.3in]{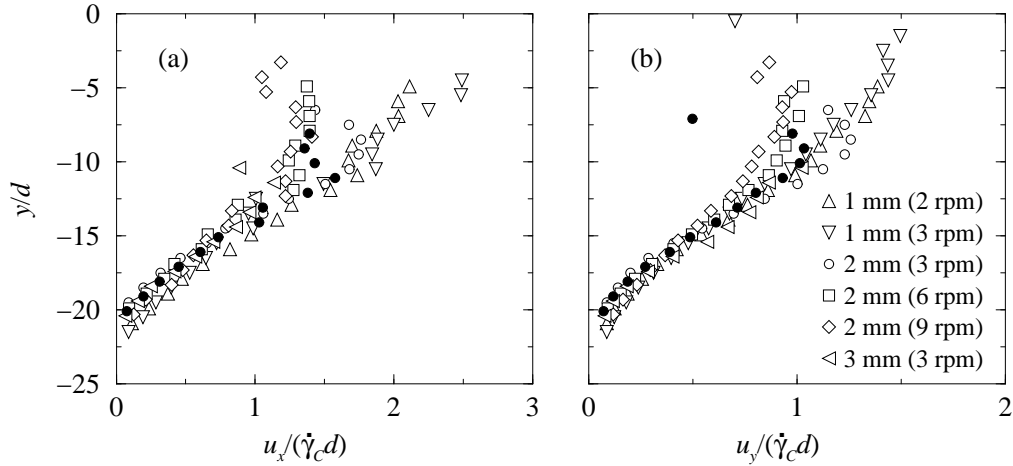}}
\caption{Scaled profiles of the components of r.m.s.\ velocity in the flowing layer. Filled symbols denote data for brass particles rotated at 3 rpm.} \label{fig:txtyscal}
\end{figure}

The r.m.s.\ velocity profiles scaled using the characteristic shear rate $\dot{\gamma}_C$ and shifted by the same distance as that for the mean velocity profiles are shown in figure~\ref{fig:txtyscal}a. The scaled profiles collapse to a single curve over most of the layer except for the region near the free surface, though the relative scatter in this case is greater as compared to the mean velocity profile.

\begin{figure}
\centering {\includegraphics[width=5.3in]{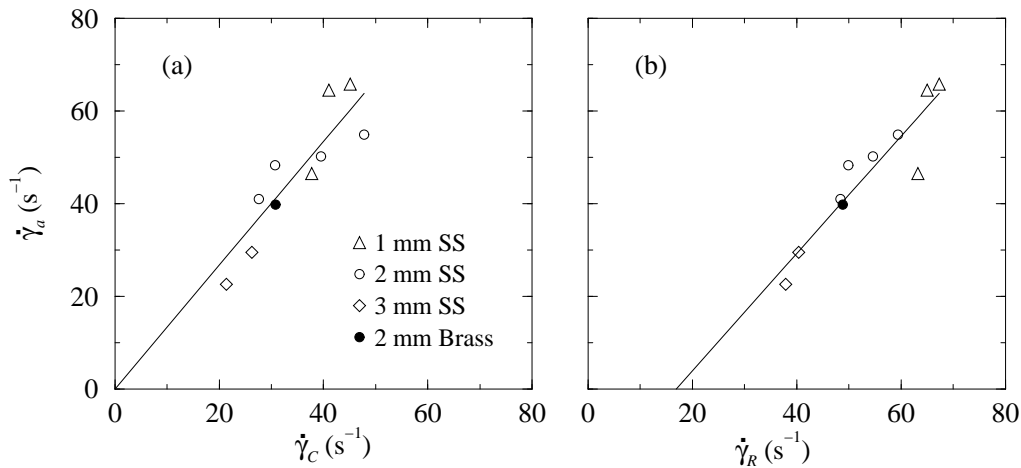}}
\caption{(a) Comparison of the measured average shear rate ($\dot{\gamma}_a$) to the shear rate predicted from \citet{kha01b} ($\dot{\gamma}_c$). (b) Comparison of the measured average shear rate ($\dot{\gamma}_a$) to the shear rate predicted from \citet{raj03} ($\dot{\gamma}_R$). Symbols show the experimental data for all the cases studied.  Solid lines show linear fits to the data.} \label{fig:prec-shear-rate}
\end{figure}

We define an average shear rate for a velocity profile, $\dot{\gamma}_a$, as the velocity gradient obtained by fitting a straight line to the linear portion of the velocity profile. An example of such a fit is shown in figure~\ref{fig:vxvsy}a. We note that the magnitude of the shear rate varies depending on the number of points included in the fitting. This variation is, however, less than $10\%$ when the number of points that are assumed to lie in the ``linear'' portion are varied. Figure~\ref{fig:prec-shear-rate} shows a comparison between the measured shear rates and the characteristic shear rate $\dot{\gamma}_C$. Although there is scatter in the data, there is reasonable agreement between the model predictions and experiment. A least squares fit gives $c=0.56$ which is different from the value estimated by \citet{orp01} ($c \approx 1.5$). The latter value was based on shear rates calculated assuming a linear velocity profile across the entire flowing layer. The velocity profiles are linear only in the central region of the layer and the shear rates are lower near the free surface and the bed. The shear rates calculated by \citet{orp01} are, thus, lower than the values corresponding to the central linear part, which results in a higher value for $c$. 

The average shear rate ($\dot{\gamma}_a$) is also proportional to the shear rate proposed by \citet{raj03} ($\dot{\gamma}_R$), as shown in figure~\ref{fig:prec-shear-rate}b. However, a linear fit gives a significant negative intercept, with $\dot{\gamma}_a=1.26(\dot{\gamma}_R-16.9)$. This comparison indicates that $\dot{\gamma}_C$ may give a better description of the flow than $\dot{\gamma}_R$.

\begin{figure}
\centering {\includegraphics[width=5.3in]{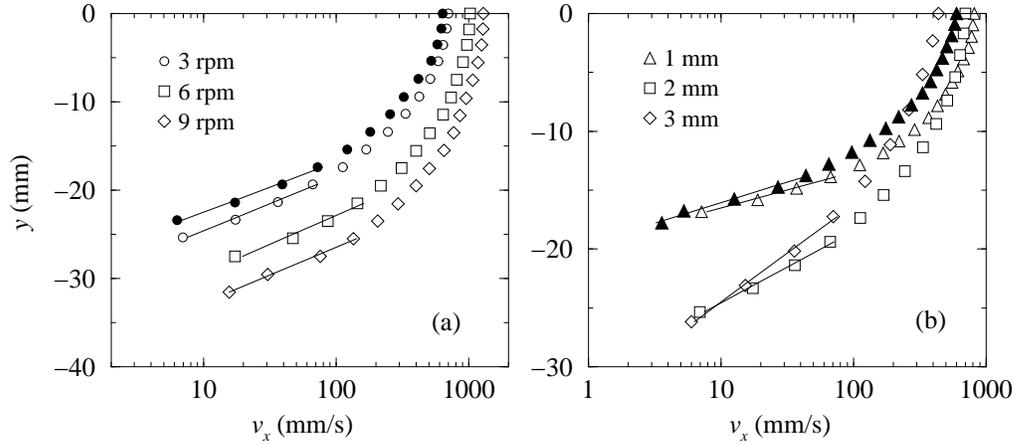}}
\caption{Mean velocity ($v_{x}$) profiles across the flowing layer plotted on a semi-log scale. Solid lines show exponential fit to the linear portion of the profiles. (a) Open symbols  denote data for $2$ mm SS balls. Filled circles denote data for $2$ mm brass particles rotated at 3 rpm. (b) Open symbols denote data for SS balls rotated at 3 rpm. Filled triangles denote data for $1$ mm SS balls  rotated at 2 rpm.} \label{fig:vxfit2}
\end{figure}

\begin{figure}
\centering{\includegraphics[width=5.3in]{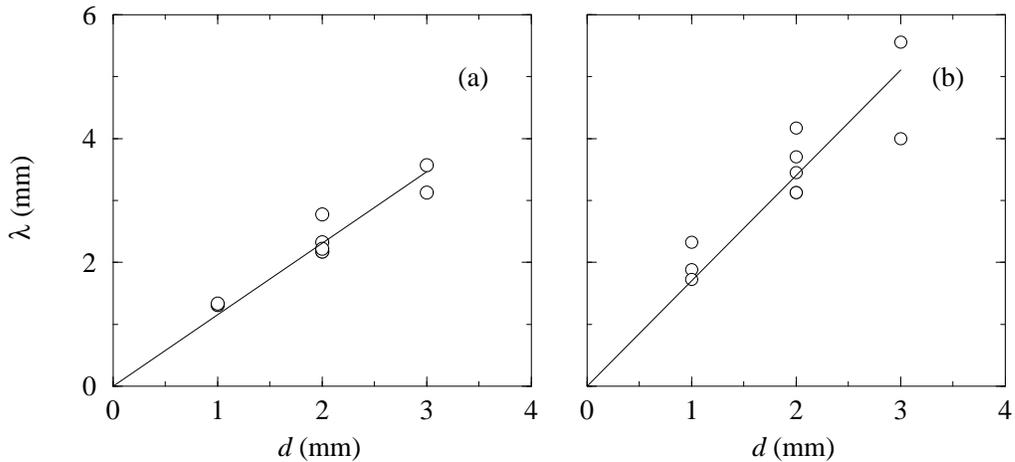}}
\caption{Variation of the characteristic  length ($y^*$) in the exponential fit with the particle diameter ($d$) for (a) mean velocity profile (b) r.m.s.\ velocity profile.} \label{fig:expcharlen}
\end{figure}

A semi-log plot of the velocity profile indicates that the exponential region of the velocity profile spans 3-4 particle diameters (figure~\ref{fig:vxfit2}). We fit straight lines to this data to obtain the characteristic decay length, $\lambda$, defined by
\begin{equation} \label{eq:expfit}
v_{x} = v_{0}e^{-y/\lambda}
\end{equation}
where $v_{0}$ is a constant of proportionality. The variation of the characteristic length $\lambda$ with the particle diameter ($d$) for the different cases studied is shown in figure~\ref{fig:expcharlen}a. The values of $\lambda$ depend on the particle size ($d$), but are nearly independent of rotational speed and material type. A linear fit to the data yields $\lambda/d=1.1\pm 0.2$. This result indicates a relatively sharp decay of the velocity with depth in the bed since the characteristic length is just a little more than one particle diameter.  A linear variation of the characteristic decay length with particle diameter was obtained by \citet{kom01} for exponential region in heap flow. A significantly larger characteristic length ($10d$) was obtained by \citet{moh97} for dense flows in a vertical channel.

An exponential decay is also found for the portion of the r.m.s.\ velocities near the base of the layer. A plot of the characteristic length with the particle diameter obtained from fits to the r.m.s.\ velocity ($u$) is shown in figure~\ref{fig:expcharlen}b. A linear fit to the data yields $\lambda/d=1.7\pm0.3$. Thus the r.m.s.\ velocity decays slightly slower with depth than the mean velocity.

\subsection{Rheology}
\label{sec:rheol}

\begin{figure}
\centering {\includegraphics[width=5.3in]{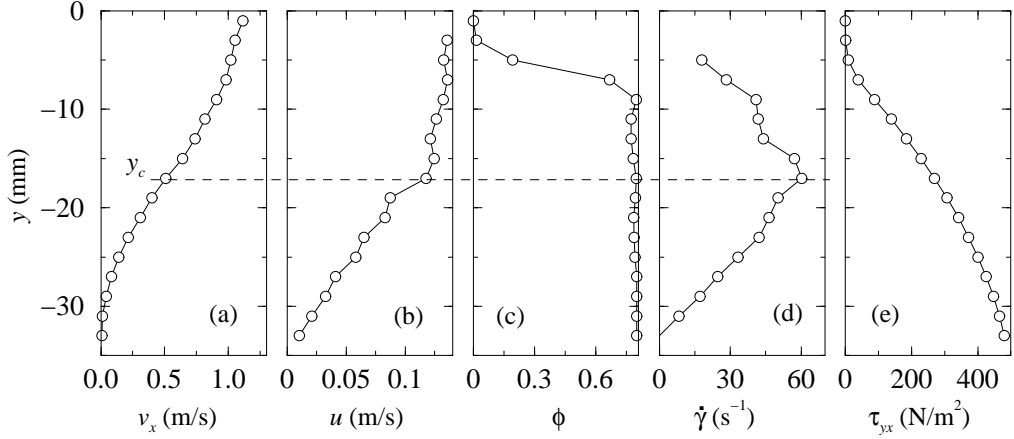}}
\caption{Profiles of the mean velocity ($v_{x}$), r.m.s.\ velocity ($u$), area fraction ($\phi$), shear rate ($\dot{\gamma}$) and shear stress ($\tau_{yx}$) in the flowing layer for $2$ mm SS balls rotated at $6$ rpm. Dashed line  denotes the transition depth ($y_c$).} \label{fig:basedata}
\end{figure}

The variation of the measured quantities with depth in the layer for one case are shown side-by-side in figure~\ref{fig:basedata} to give an overall qualitative view of the rheology. The shear rate profile, obtained by numerical differentiation of the mean velocity profile ($\dot{\gamma}(y)=dv_x/dy$), is shown in figure~\ref{fig:basedata}d. For all the cases studied, the shear rate increases to a maximum with increasing depth and then decreases as the fixed bed is approached. We note that a relatively smooth curve is obtained as a result of the averaging with the bin size equal to the particle diameter. An oscillating shear rate profile is obtained, as reported in previous works \citep{mueth00, mueth03, hill03}, if a smaller bin size is used. The fluctuations in the shear rate profile arise from the magnification of errors as a result of taking a derivative of the velocity profile data. The fluctuations are larger for higher rotational speeds and smaller particles. The results indicate that the velocity profile is not linear anywhere in the flowing layer, although the region around the maximum may be approximated to be linear. The shear rate decreases to a small value near the free surface. 

The graphs in figure~\ref{fig:basedata} indicate that on the basis of rheological behaviour, the layer can be considered to comprise two  regions with different flow characteristics. Above a critical depth, $y_c$, corresponding to the depth of the shear rate maximum, the shear stress increases with shear rate. However, in the region below $y_c$, the trend is reversed and the shear stress increases while the shear rate decreases. There also appears to be a transition in the r.m.s.\ velocity at $y_c$: above $y_c$ the profile is nearly flat, whereas it decreases with depth below $y_c$. Similar results are obtained for all cases studied. 

\begin{figure}
\centering {\includegraphics[width=5.3in]{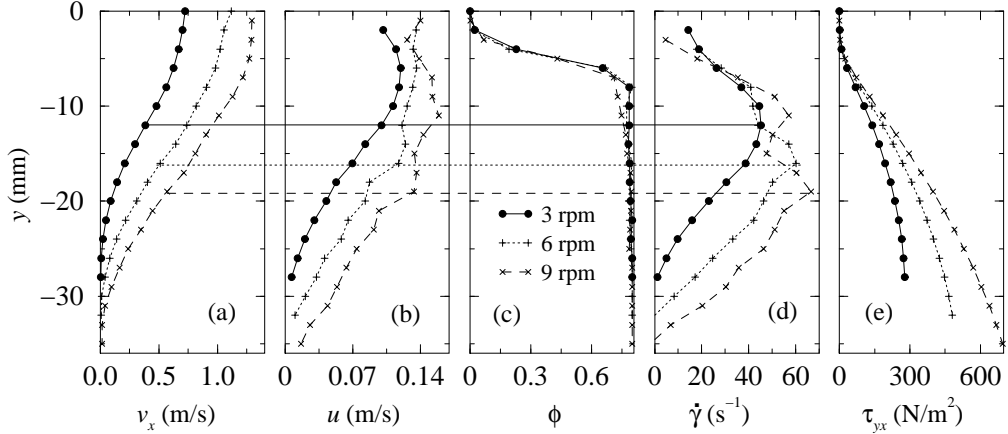}}
\caption{Profiles of the mean velocity ($v_{x}$), r.m.s.\ velocity ($u$), area fraction ($\phi$), shear rate ($\dot{\gamma}$) and shear stress ($\tau_{yx}$) in the flowing layer for $2$ mm SS balls at different rotational speeds. Horizontal lines denote the transition depth for each rotational speed.} \label{fig:basedatarpm}
\end{figure}

\begin{figure}
\centering {\includegraphics[width=5.3in]{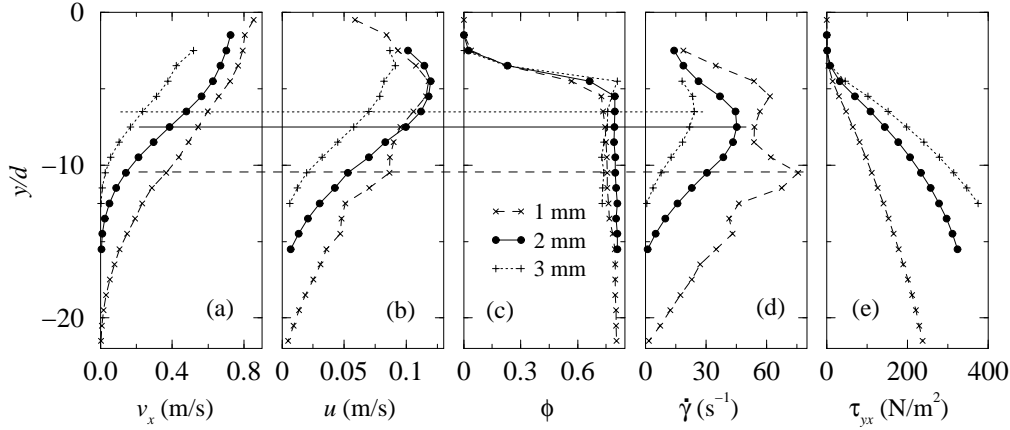}}
\caption{Profiles of the mean velocity ($v_{x}$), r.m.s.\ velocity ($u$), area fraction ($\phi$), shear rate ($\dot{\gamma}$) and shear stress ($\tau_{yx}$) in the flowing layer for different diameter SS balls at 3 rpm. Horizontal lines denote the transition depth for each particle size.}  \label{fig:basedatasize}
\end{figure}

\begin{figure}
\centering {\includegraphics[width=5.3in]{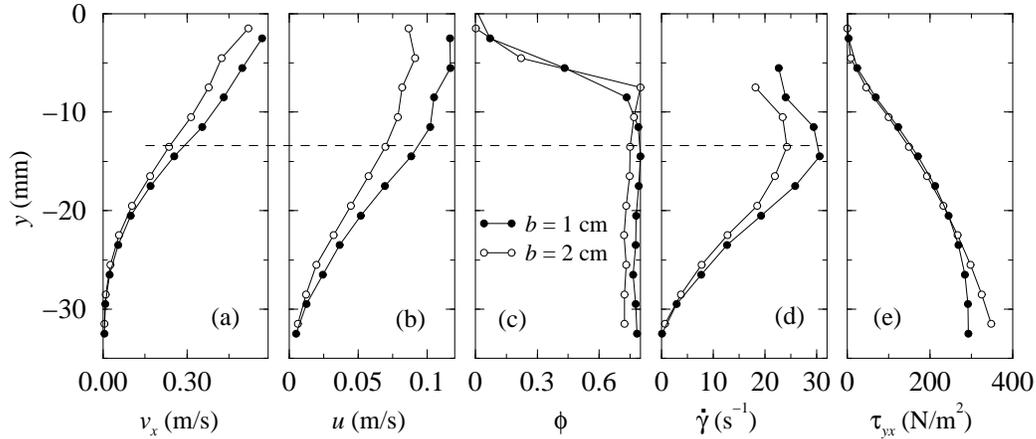}}
\caption{Profiles of the mean velocity ($v_{x}$), r.m.s.\ velocity ($u$), area fraction ($\phi$), shear rate ($\dot{\gamma}$) and shear stress ($\tau_{yx}$) across the flowing layer for $3$ mm SS balls at 3 rpm for different gap widths ($b$). The dashed line denotes the transition depth.}  \label{fig:basedatab}
\end{figure}

Figure~\ref{fig:basedatarpm} shows a comparison of profiles for 2 mm particles and different cylinder rotational speeds. All graphs for a particular rotational speed are shifted along the $y$ axis so that the area fraction profiles superimpose on each other. This ensures that the free surface in all cases is located at the same $y$ position so that depths may be compared. The transition in the flow is seen in all the cases and the data indicate that $y_c$ increases with rotational speed. Figure~\ref{fig:basedatasize} shows a comparison of profiles for different diameter particles rotated at 3 rpm. In this case the depth is rescaled by the particle diameter before shifting the graphs to superimpose the area fraction profiles. The results indicate that the scaled depth decreases with increasing particle diameter. Finally, figure~\ref{fig:basedatab} shows data for 3 mm particles for two different gap widths ($b$). Although the mean velocity and r.m.s.\ velocity profiles are slightly different for the two gaps, the transition depth does not depend on the gap width.

\begin{figure}
\centering{\includegraphics[width=5.3in]{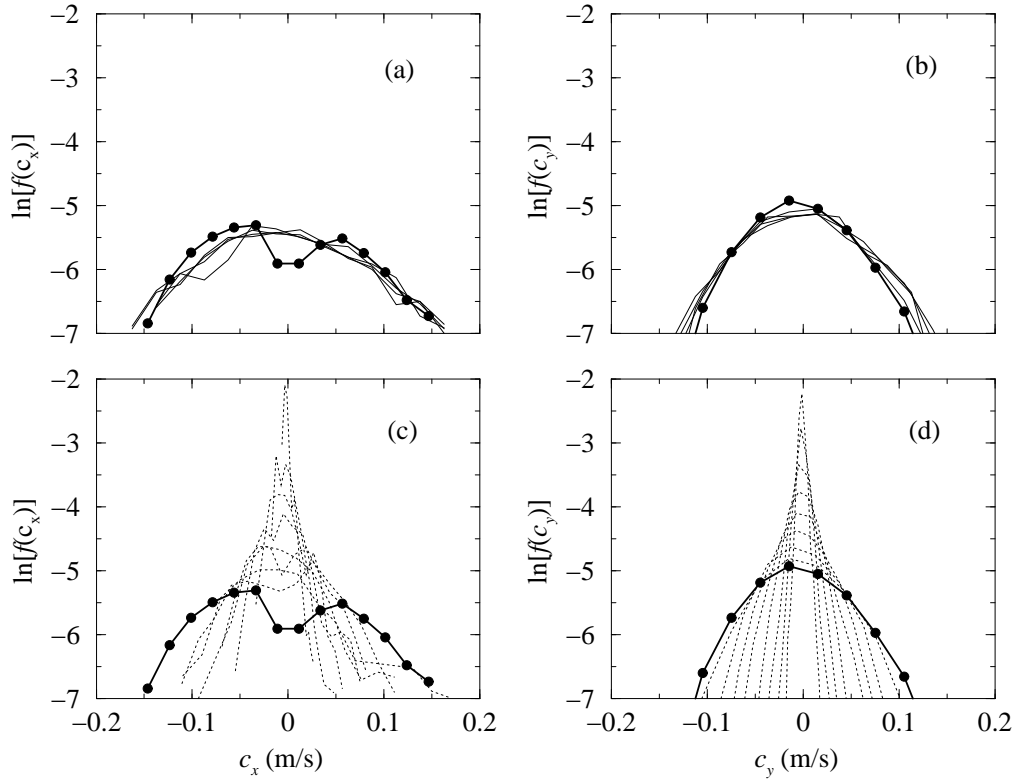}}
\caption{Velocity distributions at different depths the layer for $2$ mm SS balls rotated at $3$ rpm. Thick solid line with filled circles represents the data at the transition point. (a),(b) Region at and above the transition point. (c),(d) Region at and below the transition point.} \label{fig:dist_layer}
\end{figure}

Evidence for transition in the flow is also seen in the velocity distributions. Figure~\ref{fig:dist_layer} shows the velocity distributions at different locations in the layer for 2 mm particles rotated at 3 rpm. The distributions vary significantly with depth in the layer.  The $c_{y}$ distribution is Gaussian for all points above the transition point ($y\geq y_c$) (figure~\ref{fig:dist_layer}b). Below the transition point ($y<y_c$) it gradually evolves to a Poisson distribution with increasing depth in the layer (figure~\ref{fig:dist_layer}d). The $c_x$ distribution is Gaussian above the transition point (figure~\ref{fig:dist_layer}a). However, at and below the transition point the behaviour is complex and bimodal distributions are obtained at varying locations in the layer (figure~\ref{fig:dist_layer}c). Similar behaviour is obtained for all the cases studied. The $c_{y}$ distributions below the transition point are qualitatively similar to the distributions obtained by \citet{mueth03} for dense Couette flow.

\begin{figure}
\centering{\includegraphics[width=5.3in]{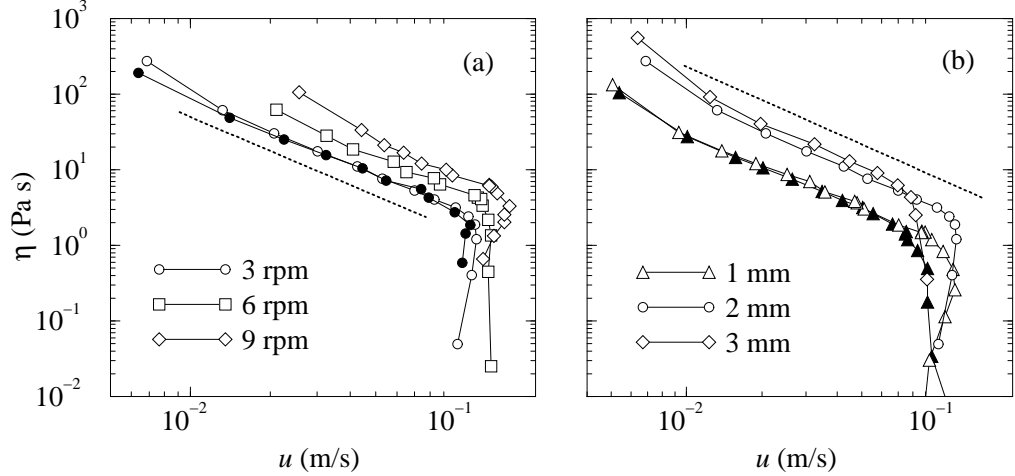}}
\caption{Apparent viscosity $\eta$ versus r.m.s.\ velocity $u$ for the different cases studied. (a) Open  symbols represent experimental data for $2$ mm SS balls. Filled circles represent data for $2$ mm brass particles rotated at $3$ rpm. (b) Open symbols represent experimental data for the SS balls rotated at $3$ rpm. Filled triangles represent data for $1$ mm SS balls rotated at $2$ rpm. Dotted lines represent a least squares fit (slope $\approx -1.5$) to the data in the linear region.} \label{fig:visctemp}
\end{figure}

In the region below the transition depth ($y_c$), the decrease in shear rate with depth, accompanied by an increase in shear stress with depth, implies a sharp increase in the apparent viscosity, defined as $\eta = \tau_{yx}/\dot{\gamma}$. Figure~\ref{fig:visctemp} shows the variation of the apparent viscosity ($\eta$) with the r.m.s.\ velocity ($u$). The granular temperature is $T=u^2$, thus figure~\ref{fig:visctemp} may be considered to be analogous to the variation of viscosity with temperature in molecular fluids. The viscosity varies by more than three orders of magnitude and all the data show qualitatively similar behaviour. Near the free surface the viscosity is low and increases at constant $u$, essentially due to the increase in the volume fraction of solids. Below the transition point there is a gradual increase in viscosity with depth corresponding to a decreasing $u$. The variation of $\eta$ with $u$ in the region just below the transition in all cases is power law with an exponent of about $-1.5$. At lower values of $u$ the curves turn upward and rate of increase of $\eta$ appears to become higher.

\begin{figure}
\centering {\includegraphics[width=5.3in]{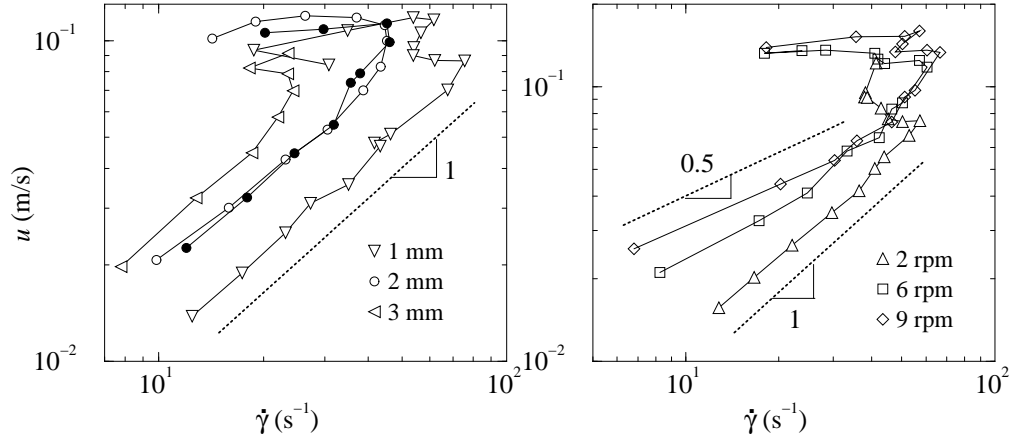}}
\caption{Variation of the r.m.s.\ velocity ($u$) with the shear rate ($\dot{\gamma}$) in the the flowing layer. Dotted lines show least squares fits to the linear portions of the graphs. (a) Particles of different diameter and a rotational speed of 3 rpm.  Filled symbols denote data for $2$ mm brass particles. (b) 2 mm particles at different rotational speeds.} \label{fig:tygam}
\end{figure}

Figure~\ref{fig:tygam} shows the variation of the r.m.s.\ velocity ($u$) with shear rate ($\dot{\gamma}$). There is a power law dependence between the two in the region below the transition corresponding the high density region. The exponent ($\alpha$) is independent of the particle diameter (figure~\ref{fig:tygam}a), but appears to decrease systematically with increasing rpm, ranging from $\alpha=1$ for low rotational speeds to $\alpha=0.5$ for 9 rpm. However, data over a wider range are required to verify these results. Power law variation between the r.m.s. velocity and the shear rate have been obtained previously by \citet{boc01} and \citet{mueth03} for dense Couette flows. The value of the exponent reported by them is around $\alpha=0.5$. \citet{azanza99} found $\alpha=1$ for rapid flows.

The above results indicate the presence of two regimes in the flowing layer with a sharp transition point. In the upper region near the free surface, the behaviour is fluid-like and the velocity distributions are Maxwellian. Below the transition point the material appears to be an amorphous soft solid, increasing in strength with depth in the layer. The transition to this solid-like regime occurs at a relatively large r.m.s.\ velocity ($\approx 0.1$ m/s). Qualitatively similar behaviour is obtained for different particles and operating speeds.

We conjecture that the sharp transition occurs because of the formation of a percolated network of particles in extended contact with each other. This is in contrast to the fluid-like regime where the particles interact through collisions. The contact network coexists with fluid-like domains and the fraction of particles which are part of the network increase with depth. This picture for the region below the transition point is broadly consistent with recent measurements of \citet{bon02b} in which the flowing clusters were identified, as well as with non-local models based on the coexistence of particle chains and fluid-like material \citep{mills99,bon03}. 

Table~\ref{tab:transdepth} gives the thickness of the fluid-like ($\delta_f$) and the solid-like ($\delta_s$) regions obtained from the data in figures \ref{fig:basedatarpm} and \ref{fig:basedatasize}. The data indicates that the thickness of the fluid-like layer is independent of the particle diameter but depends on the rotational speed (i.e., the local flow rate, $q$). In contrast, the thickness of the solid-like region is independent of rpm but proportional to the particle diameter. The latter result is consistent with the finding that the characteristic length for exponential decay of the velocity ($\lambda$) is proportional to diameter but independent of rotational speed.

\begin{table}
\centerline{\begin{tabular}{cccc}
$d$&$\omega$&$\delta_f$ &$\delta_s$\\
(mm)&(rpm)& (mm) &(mm)\\
1 &3 &6 &11\\
2 &3 &7 &16\\
2 &6 &11 &16\\
2 &9 &14 &15\\
3 &3 &7 &20
\end{tabular}}
\caption{Measured thicknesses of the fluid-like ($\delta_f$) and solid-like ($\delta_s$) layers for different cases.\label{tab:transdepth} }
\end{table}

\subsection[Comtheory]{Comparison to theory}
\label{sec:comparison-theory}

\citet{bon03} proposed a non-local model for the surface flow of granular material, based on an earlier work by \citet{mills99}. The model assumes the flowing granular medium as a network of transient particle chains, immersed in an assembly of particles behaving as a viscous Newtonian fluid. The stresses are then propagated through these chains in a non-local manner. The total stress is expressed as a linear combination of Coulombic frictional stresses, viscous stresses and the stresses due to the embedded chains.

The model equations, simplified for the case of surface flows in rotating cylinders for low rotational speeds ($\omega$), when inertial stresses can be neglected, yield an analytical expression for the velocity ($v_{x}$) profile (\citet{bon03})
\begin{subeqnarray} \label{eq:bonvel}
  v_{x}(y) & \simeq &\left(\frac{\lambda}{2}\right)\dot{\gamma}e^{2\delta/\lambda}e^{2y/\lambda}
  - \left(\frac{\omega}{2R}\right)(y^{2}+R^{2}) \;\;\;\;\;\;\;\; y \in
  (-R,-\delta)  \\
  & \simeq & \dot{\gamma}\left(y+\delta+\frac{\lambda}{2}\right) -
  \left(\frac{\omega}{2R}\right)(y^{2}+R^{2}) \;\;\;\;\;\;\;\; y \in
  (-\delta,0)
\end{subeqnarray}
where $\lambda$ is a function of the characteristic correlation length of the transient chains. The shear rate ($\dot{\gamma}$) and the characteristic length ($\lambda$) are parameters of the model. Using the mass balance equation at the centre of the cylinder, gives the following expression for the layer thickness
\begin{equation} \label{eq:bonmass}
  \delta  \simeq R\left(\frac{4\omega}{3\dot{\gamma}}\right)^{1/2}
\end{equation}
Equations~(\ref{eq:bonvel}) and (\ref{eq:bonmass}) are used to calculate the velocity profiles across the flowing layer. The shear rate in each case is taken to be the value obtained by fitting a straight line to the linear part of the experimental profile ($\dot{\gamma}_a$, figure~\ref{fig:prec-shear-rate}). The values of the correlation length are found to be related to the particle diameter as $\lambda/d= 4$ for all the cases.

\begin{figure}
\centering {\includegraphics[width=5.3in]{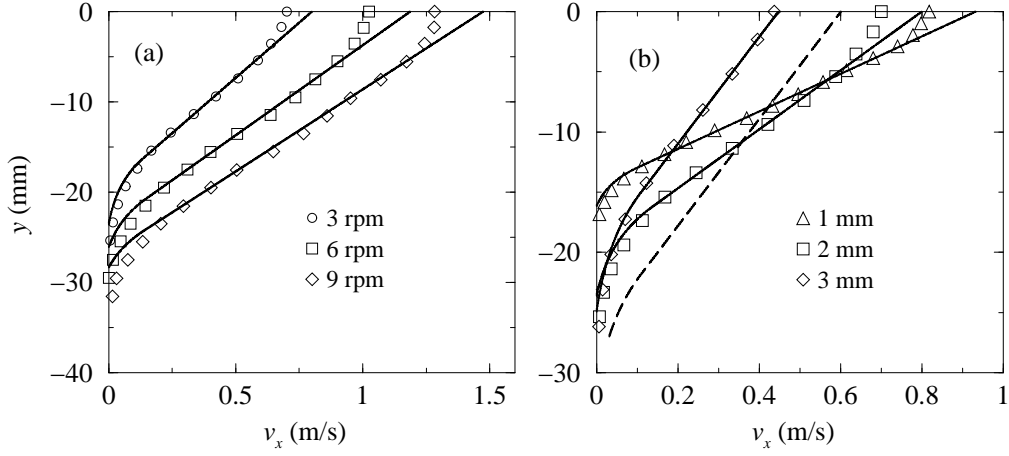}}
\caption{Mean velocity ($v_{x}$) profiles in the flowing layer. Solid lines show the predictions of the model of \citet{bon03}. (a) Symbols denote data for $2$ mm SS balls. (b) Symbols denote data for SS balls rotated at 3 rpm. } \label{fig:vxmod2}
\end{figure}

The predicted and the experimental profiles are shown in figure~\ref{fig:vxmod2}. The model predicts the exponential and the linear part of the velocity very well for all rotational speeds and particle sizes, but does not predict the flattening of the profile near the free surface. For the case of of $3$ mm balls, the layer thickness calculated from (\ref{eq:bonmass}) overpredicts the experimental layer thickness resulting in the shift of the predicted profile (dashed line in figure~\ref{fig:vxmod2}b). The reason for this mismatch is the experimental difficulty in correctly identifying the location of the free surface. If, for this case we use the layer thickness value obtained from the experimental velocity profile, a good fit is obtained.

\citet{aran02} proposed an order parameter description for the fluidization transition in granular flows. The order parameter specifies the fraction of the solid and fluid parts of the stress tensor, and is related to the number of particles in contact with a given particle on the average (coordination number). The dynamics of the order parameter ($p$) are assumed to be governed by the Ginzburg-Landau equation which reduces in the case of a steady, fully-developed unidirectional flow to
\begin{equation}  \label{eq:aran1}
0=D\frac{d^2p}{dy^2} - F(p,\mu)
\end{equation}
where $D$ is the characteristic diffusion coefficient for $p$. $F(p,\mu)$ is the derivative of the potential energy density and is given by 
\begin{equation} \label{eq:aran2}
  F(p,\mu) = (p-1)(p^{2}-2p_{*}p+p_{*}^{2}\exp[-A(\mu^{2}-\mu_{*}^{2})])
\end{equation}
where $\mu = \tau_{yx}/\tau_{yy}$ and the quantities with asterisks are model parameters. \citet*{volf03b} found $p_{*}=0.6$, $A=25$ and $\mu_{*}=0.26$ from two-dimensional discrete element simulations for $D=2$. The fluid ($\tau_{yx}^{f}$) part of the total shear stress ($\tau_{yx}$) was obtained from simulations in terms of the order parameter as
\begin{equation}  \label{eq:aran3}
  \tau_{yx}^{f} = (1-p)^{2.5} \tau_{yx} = \eta \frac{dv_{x}}{dy} 
\end{equation}
where $\eta$ is the viscosity. Simulations for 2d systems indicate $\eta=12$. In the above equations (\ref{eq:aran1} - \ref{eq:aran3}), all quantities are made dimensionless using appropriate combinations of $m$, $g$ and $d$.

We obtain the profile of the order parameter in the layer by numerical solution of (\ref{eq:aran1}) using a shooting method with the experimentally obtained profile for the local friction coefficient ($\mu(y)$) as an input. The free surface is assumed to be a no flux boundary for the order parameter ($Ddp/dy=0$ at $y=0$) and a pure solid ($p=1$) is assumed far from the free surface ($y=\delta_A>>1$). The profiles are independent of the domain depth ($\delta_A$) when it is large enough. The shear rate ($\dot{\gamma}=dv_x/dy$) profile is then calculated from (\ref{eq:aran3}a) using the experimentally obtained shear stress profile. 

\begin{figure}
\centering {\includegraphics[width=5.3in]{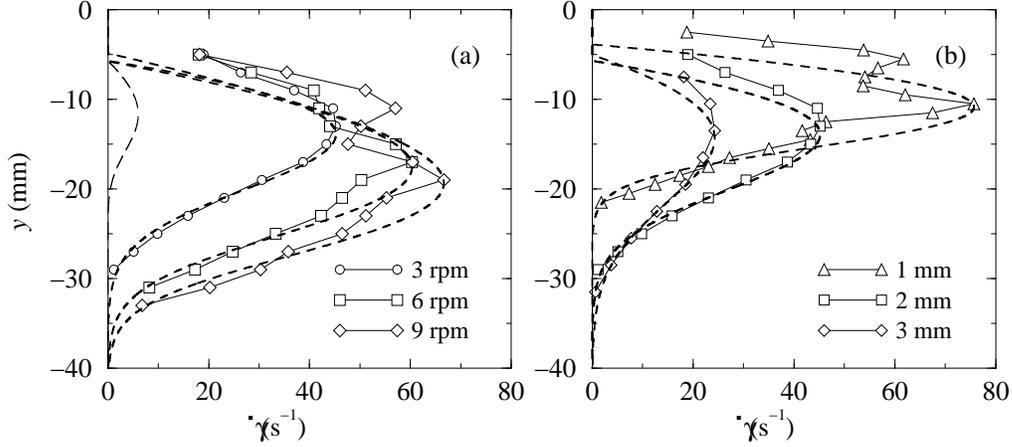}}
\caption{Comparison of the shear rate profiles predicted by the model of \citet{aran02} (dashed lines) to the measured shear profiles. (a) Symbols denote data for $2$ mm SS balls. (b) Symbols denote data for SS balls rotated at 3 rpm. } \label{fig:vxmod1}
\end{figure}

\begin{figure}
\centering {\includegraphics[width=3in]{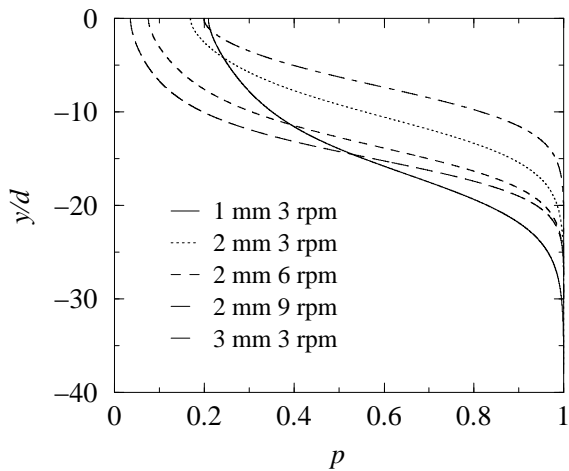}}
\caption{Variation of the order paramter ($p$) with depth in the flowing layer corresponding to the cases described in table~\ref{tab:model-fits1}.} \label{fig:aransonp}
\end{figure}

The computed shear rate profiles are compared to experimentally measured profiles in figure~\ref{fig:vxmod1}. This is a more stringent test as compared to matching the model to the measured velocity profiles. Predictions of the shear rate profile using the parameters from 2d simulations give qualitative agreement with experimental profiles. One such calculation for 2 mm particles rotated at 3 rpm is shown in figure~\ref{fig:vxmod1} (long dashed line) using the values given above for $A$, $D$ and $\eta$, but taking $\mu*=0.4$. The latter was necessary since no solution satisfying the boundary conditions was obtained for $\mu*=0.26$ for the particular experimental profile for $\mu(y)$ used. We found excellent quantitative agreement between the model and experimental results using $A=5$ and $D=2$ for four of the cases and $A=10$ and $D=2$ for one case, as shown in the figure~\ref{fig:vxmod1}. However, different values of $\mu*$ and $\eta$ are required to get a good fit and these are given in table~\ref{tab:model-fits1}. We found that $\mu*$ essentially determines the depth of the layer and thus $\mu*$ was adjusted to match the experimental and theoretical depths. The viscosity ($\eta$) is a proportionality factor and we calculated it as the ratio of the maximum computed shear rate to the maximum experimental shear rate for each profile. Once $A$ and $D$ are fixed, the calculated values of $\mu*$ and $\eta$ are thus uniquely determined. The parameter $A$ did not significantly affect the shape of the profiles when varied over a significant range ($A\in(5,25)$), but determined the range of $\mu*$ for which solutions satisfying the boundary conditions could be obtained. Similarly, changing $D$ over narrow ranges close to $D=2$ did not change the profiles significantly, but strongly affected the range of $\mu*$ for which solutions could be obtained. The graph of the order parameter variation with depth for all the cases is shown in figure~\ref{fig:aransonp}. Note that according to the model, the material is not a pure fluid ($p>0$) at the free surface ($y=0$).

The values of the parameter $\mu*$ correlate well with the effective friction coefficient ($\mu_i$), also given in table~\ref{tab:model-fits1}. Thus $\mu*$ depends on the frictional characteristics of the material as may be expected. The viscosity is nearly constant for different particle sizes and increases with rotational speed. The latter may be expected given the increase in the granular temperature with rotational speed.

\begin{table}
\centering
\begin{tabular}{ccccccc} 
    Material & $d$ & $\omega$ &$A$ & $\mu*$ & $\eta$ &$\mu_i$ \\ 
    & (mm) & (rpm) & &  & \\ \hline
    & $1$ & $3$ &5& 0.36 & 0.97 &0.36\\
    & $2$ & $3$ &5& 0.40 & 0.85 &0.34\\
    SS & $2$ & $6$ &5& 0.57 &1.74 &0.52\\
    & $2$ & $9$ &5& 0.81 &2.77 &0.77\\
    & $3$ & $3$ &10& 0.41 & 0.84 &0.37\\ 
\end{tabular}
\caption{Values of fitting constants for the model of \citet{aran02}. The values of the effective coefficient of friction at the base of the layer, $\mu_i=\mu(y_s)$, are also given.} \label{tab:model-fits1}
\end{table}

\begin{figure}
\centering {\includegraphics[width=5.3in]{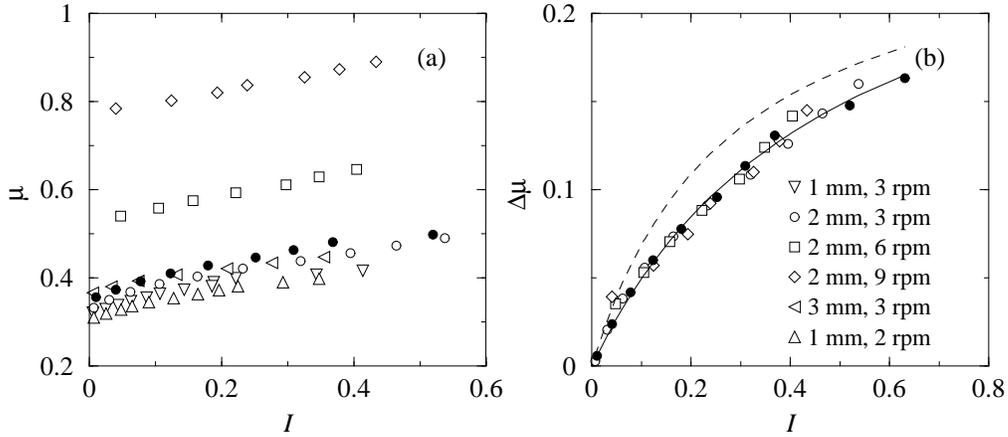}}
\caption{(a) Variation of the effective friction coefficient ($\mu$) with scaled shear rate ($I$) for different particle sizes and cylinder rotational speeds as indicated in the legend. Filled symbols correspond to 2 mm brass balls and a rotational speed of 3 rpm. (b) Increase in coefficient of friction ($\Delta \mu$) with scaled shear rate ($I$) for 2 mm particles. The solid line is a fit of (\ref{eq:pouliq}) and the dashed line gives predictions of the same equation but using the parameters obtained by \citet{jop05}.} \label{fig:poul2}
\end{figure}

Finally, we compare the experimental results obtained to the rheological model of Pouliquen (\ref{eq:pouliq}). A direct comparison is possible in this case since $\mu$ and $I$ can be estimated from experimental data. Figure~\ref{fig:poul2}a shows the variation of the local coefficient of friction ($\mu$) with the scaled shear rate ($I$) for the different data sets. The data sets are truncated and only the points in the dense flow region below the shear rate maximum are shown. The data for the lower rotational speeds are all clustered toegther while the values of $\mu$ for the higher rotational speeds are significantly higher. This follows from the higher values of $\mu_i$ for the higher rpm data seen in figure~\ref{fig:del_betam}. We take $\mu_s$ to the value of $\mu(I)$ extrapolated to $I=0$. The range of $\mu_s$ for the low rpm data corresponds to angles ranging from 17 to 20 deg., which reflects the error in estimation of $\mu$. The values are in reasonable agreement with the results of \citet{jop05}. Figure~\ref{fig:poul2}b shows the variation of $(\mu-\mu_s)$ with the scaled shear rate ($I$) for 2 mm particles and for different cyinder rotational speeds (3, 6 and 9 rpm). The collapse of data is reasonable, and the solid line is a fit to all the sets combined. The fit yields $\Delta\mu=(\mu_2-\mu_s)=0.29$ and $I_0=0.49$. The values reported by \citet{jop05} are $\Delta\mu=0.261$ and $I_0=0.279$, and a plot of the model using these values is shown in figure~\ref{fig:poul2}b (dashed line). The difference between the two is not very large considering the the different materials used here. Fits of the model gave slightly different model parameter values for 1 mm particles ($\Delta\mu=0.18$, $I_0=0.48$) and 3 mm particles ($\Delta\mu=0.20$, $I_0=0.43$), but only one set of data was available for fitting in each case.

\section{Conclusions}
\label{sec:conclusions}

An experimental study of surface flow in quasi-2d rotating cylinders is carried out using streakline photography. Experiments are performed for a wide range of rotational speeds and particle sizes. The mean velocity profiles are linear over most of the layer depth for all the cases with an exponential decay near the base of the flowing layer and a flattened region near the free surface.  The r.m.s.\ velocities are nearly constant in a region near the free surface followed by a linear decrease over part of the layer and an exponential decay into the bed.  Further, the r.m.s.\ velocities show a distinct anisotropy, with the component in the flow direction larger by a factor of about $1.5$ than that perpendicular to the flow direction. The number densities are nearly constant across the flowing layer with a sharp decrease near the free surface. The shear and the normal stresses increase linearly with the depth of the flowing layer initially, but the shear stress deviates from linearity due to wall friction. Streaming stresses are found to be significant in the region near the free surface. The experimental velocity profiles reported here are broadly similar to those in previous works, however, a wider range of system parameters is explored in this work.

The experimental data is analyzed in several ways. A simple scaling is found to work very well. The mean velocity profiles for different cases scaled using the shear rate from \citet{orp01} and the particle diameter collapse to a single curve. A similar scaling works well for the r.m.s.\ velocity except for the region near the free surface. The shear rate obtained from a least square fit of the linear portion of the mean velocity profile is reasonably predicted by the model shear rate (\ref{eq:shear}) but with a different value for the empirical constant ($c$). The mean velocity and r.m.s.\ velocity profiles near the base of the layer are well described by an exponential function of the form $\exp(-y/\lambda)$, and the characteristic decay length ($\lambda$) is proportional to the particle diameter ($d$) in both cases, with $\lambda \approx 1.1d$ for the mean velocity and $\lambda \approx 1.7 d$ for the r.m.s.\ velocity. Thus the r.m.s.\ velocity decays more slowly than the mean velocity.

The shear rate profile obtained by numerically differentiating the mean velocity profile data indicates that the shear rate is not constant anywhere in the layer but increases to a maximum and then decreases to zero with depth in the layer. A transition in the r.m.s.\ profile occurs at the same depth as the shear rate maximum: above the transition depth the r.m.s.\ velocity is nearly constant whereas below the transition depth it decreases with depth. Evidence of a transition is also seen in the variation of the velocity distributions with depth: Above the transition point the distributions are Gaussian and below the transition point the velocity distributions gradually approach a Poisson distribution. 

The rheology of the flow is characterized in terms of the variation of the apparent viscosity with r.m.s.\ velocity. The curves indicate a relatively sharp transition at the shear rate maximum, and in the region below this point the apparent viscosity varies as $\eta\sim u^{-1.5}$. The transition observed in the measurements appears to be a percolation type of transition, in which there is an increasing fraction of particles below the transition depth which are confined, due to formation of a network of particles in extended contact with each other. From the viewpoint of rheology, the flow comprises two layers: an upper low viscosity layer with a nearly constant r.m.s.\ velocity and a lower layer of increasing viscosity with a decreasing r.m.s. velocity. The thickness of the upper layer depends on the local flow rate but is independent of the particle diameter. In contrast, the lower layer thickness depends of the particle diameter but is independent of flow rate. Thus after establishment of the lower layer any further increase in the local flow rate results in the material passing through the upper layer. The rheological characteristics of the two layers are different.
 
Predictions of both models for the mean velocity match the experimental velocity profiles reasonably well. The model of \citet{bon03} requires only two parameters : the shear rate, which can be independently obtained and the characteristic length which is found to be related in a simple way to the particle diameter ($\lambda= 4d$). This model, however, does not describe the low density region near the free surface. The model of \citet{aran02} describes the entire velocity profile, however, experimental methods for independently estimating the parameter values need to be devised. A comparison of the predictions of the rheological model proposed by Pouliquen \citep{jop05} to experimental data also shows reasonable agreement when the increase in friction coeffcient with shear rate is considered. However, the coefficient of friction is found to be significantly higher at high local flow rates (corresponding to high rotational speeds) which is not in agreement with the model. The models considered show promise for prediction of the mean velocity profile and effective friction coefficient. More detailed rheological models are required for predicting r.m.s.\ velocities and velocity distributions which are important for transport.

The results presented in this paper give a reasonably detailed picture of the rheology of surface granular flows in quasi-two-dimensional systems. Although the results may be expected to to valid in qualitative terms for 3d systems, wall effects cause significant quantitative differences. The walls impose two effects: one is side wall friction, which is accounted for approximately in this work, and the second is ordering by the steric effects of the side walls. The measurements of \citet{cou05} indicate that wall friction results in the velocity at the side wall being about 20-30\% lower than that at the middle of the gap. Our preliminary results indicate that the effects of ordering may be larger: the mean velocity in long cylinders is about half of that for a comparable quasi-two-dimensional system. Thus the ordering appears to reduce the effective viscosity of the system. Modelling such wall effects remains a challenge.

\begin{acknowledgments}
We are grateful for discussions with J. T. Jenkins, O. Pouliquen and I. Aranson. The computer program for the calculations in Figs. 29 and 30 was provided by I. Aranson, and we are grateful for his help. This paper was written while one of us (DVK) was visiting Institut Henri Poincar\'{e} and the financial support of CNRS during the stay is gratefully acknowledged.
\end{acknowledgments} 

\bibliography{ref}
\end{document}